\RequirePackage{fix-cm}
\documentclass[smallextended]{svjour3}
\smartqed  
\usepackage{graphicx}
\usepackage{mathtools}
\newcommand{\Mcore}{M_{\rm core}}
\begin{document}

\title{Uranus and Neptune: Origin, Evolution and Internal Structure}

\author{Ravit Helled         \and
        Nadine Nettelmann \and 
        Tristan Guillot }


\institute{R. Helled \at
              Institute for Computational Science \\
              Center for Theoretical Astrophysics \& Cosmology \\
              University of Zurich\\
              Winterthurerstr. 190, CH-8057 Zurich
              Switzerland\\
              Tel.: +41(0)446356189\\
              \email{rhelled@physik.uzh.ch}           
           \and
           N. Nettelmann \at
              Institute for Planetary Research, German Aerospace Center (DLR)\\ 
              12489, Berlin, Germany\\\\
              T. Guillot \at
              C\^ote d'Azur, Lagrange, Observatoire de la C\^ote d'Azur, CNRS \\ 
              Nice Cedex 4, France
}

\date{Received: date / Accepted: date}
\maketitle

\begin{abstract}
There are still many open questions regarding the nature of Uranus and Neptune, the outermost planets in the Solar System. 
In this review we summarize the current-knowledge about Uranus and Neptune with a focus on their composition and internal structure, formation including potential subsequent giant impacts, and thermal evolution. 
We present key open questions and discuss the uncertainty
in the internal structures of the planets due to the possibility of non-adiabatic and inhomogeneous interiors. 
We also provide the reasoning for improved observational constraints on 
their fundamental physical parameters such as their gravitational and magnetic fields, rotation rates, and deep 
atmospheric composition and temperature. 
Only this way will we be able to improve our understating of these planetary objects, and the many similar-sized objects orbiting  other stars. 
\keywords{ Planets and satellites: formation --
                Planets and satellites: interiors --
                Planets and satellites: ice planets --
                Planets and satellites: composition --
                Planets and satellites: individual: Uranus, Neptune}
\end{abstract}

\newpage
\section{Introduction}
\label{intro}
Uranus and Neptune have masses of about 14.5 and 17 M$_{\oplus}$ (Earth mass), respectively. Their sizes are about four times that of the Earth, and they are located at 20 and 30 AU from the Sun.    
Uranus and Neptune represent a unique class of planets in our Solar System, and are often referred as the "ice giants". They are different from the terrestrial planets since they are significantly more massive, consist mostly of volatile materials (ice-forming elements such as oxygen and carbon), and have much colder atmospheres. They also differ from Jupiter and Saturn, the gas giants, since they are significantly smaller and their compositions are not dominated by hydrogen-helium (H-He).

Despite the similar masses and radii of Uranus and Neptune, there are also noticeable differences between these planets, such 
as their atmospheric enrichment, obliquity, and thermal emission. 
Uranus' radius is larger than Neptune's but its mass is smaller, making Neptune denser than Uranus by $\sim$30\%. 
Their inferred moment of inertia (MoI) values from interior models suggest that Uranus is more centrally condensed than Neptune.  
A distinct feature of Uranus is its large axial tilt  
 and its regular satellites, suggesting they formed from a circumplanetary disk, while 
Neptune's largest moon, Triton,  in a retrograde orbit, and was probably captured. 
In addition, Neptune's measured heat flux implies that it is still cooling, while Uranus is near equilibrium with solar insolation (e.g., Pearl et al. 1990; Pearl \& Conrath 1991), suggesting that Uranus' interior is not fully convective, and/or that it contains compositional gradients or thermal boundary layers that hinder convection (e.g., Nettelmann et al., 2016, Podolak et al., 2019). 

The available measurements of the fundamental physical properties of Uranus and Neptune such as mass, radius, and gravitational field can be used to constrain their interiors.  
However, 
there are still substantial uncertainties regarding their bulk compositions and internal structures, since the planets' interiors are complex and at the same time the available data are somewhat limited 
(e.g., Podolak et al., 1991; 1995, Guillot, 2005, Podolak \& Helled, 2012, Fortney \& Nettelmann, 2010, Nettelmann et al., 2013, Helled \& Guillot, 2018). 
In addition, the formation process of these planets remains a great challenge for planet formation theories, as well as their subsequent evolution. 
 Structure models suggest that Uranus and Neptune possess H-He atmospheres. Their atmospheres are expected to be
accreted from the protoplanetary disk gas as suggested by standard planet formation models (see below for details). 
However, it is very challenging to explain their exact atmospheric masses and why they have not grown further to become gas giants. 
The thermal evolution of the planets is complex due to the potential existence of sophisticated chemical and physical processes such as mixing, settling, phase separation, inhibited convection, development of composition gradients and boundary layers. 
\newline
Uranus and Neptune have not been explored from space in detail apart from the single Voyager 2 flybys, and many key questions regarding the planets'  regarding their origin, evolution, and internal structure remain unsolved. At the same time, we now know that exoplanets in the mass/radius-regime of Uranus and Neptune are common, as well as their smaller versions (often referred as "mini-Neptunes"). But we do not know their bulk compositions and internal structures, neither how heat is transferred in these planets, thus limiting our ability to determine their composition and the mechanisms of planet formation. 
Although these discovered exoplanets are significantly hotter than Uranus and Neptune, and are also likely to have suffered migration, they reveal the diversity of planets in this mass-range. 
It is therefore highly important to better understand Uranus and Neptune, which are clear priorities of the international planetary science community, and are also crucial to understand exoplanets.

\section{Basic Properties of Uranus and Neptune}
\label{sec:1}

{\bf Gravitational Coefficients ($J_{n}$)}\\
The measured gravitational field of a planet is used to constrain its internal structure. For giant planets in hydrostatic equilibrium  and no hemisphere asymmetries only the even gravitational harmonics ($J_{2n}$) are expected to exist.  
 Typically, the gravitational moments are inferred  from analysing the trajectories of spacecraft during flyby, especially when they approach the planet, and preferably in a polar orbit. 
The gravitational coefficients of Uranus and Neptune were measured during the Voyager 2 flybys. 
 More accurate determinations of the gravitational coefficients were then inferred from the precession of Uranus rings and the orbits of the satellites of the planets (e.g., Elliot \& Nicholson, 1984; Jacobson, 2009, 2014). 
The error bars associated with the measurements are rather large (compared to Jupiter and Saturn), and the harmonics are known only to fourth order (i.e., $J_2, J_4$). 
Jacobson and collaborators (Jacobson et al., 2007; Jacobson, 2009, 2014) re-determined the gravitational harmonics of the planets using Earth-based astrometry and observations acquired mostly with the Voyager spacecraft; and provided more accurate estimates for the gravitational harmonics with smaller error bars. 
The density profile of the planet in interior models is set to reproduce the measured gravitational field. While the exact composition is unknown, the constraints introduced by the gravitational harmonics can be used to narrow down the possible planetary composition and internal structure (see review by Helled 2019 and references therein for details). 
\newline

\noindent{\bf Rotation periods}\\
The rotation periods of giant planets in the Solar System are typically determined from radio and magnetic field data. 
However, the periodicities inferred from these data might not represent the planetary bulk rotation. Voyager 2 measurements of periodic variations in the radio signals and of fits to the  magnetic fields of Uranus and Neptune imply rotation periods of 17.24 h and 16.11 h, respectively. 
The periods inferred from the radio signals and magnetic fields might be equal if the radiation emanates from charged particles that are attached to the magnetic field lines, but the periods could also differ 
if the radiation originates from a local concentration of ions in a centrifugally loaded magnetosphere. 
In addition, it is unclear which parts of the planetary interiors are tied to the magnetic field lines, in particular in the case of Uranus and Neptune, which have multi-polar magnetic fields that are expected to originate from relatively shallow depths (e.g., Stanley \& Bloxham 2004, 2006). 

It was then suggested by Helled et al. (2010) that the Voyager 2 radio and magnetic periods do not represent the deep interior rotation periods, and they proposed modified rotation periods for the planets by searching for the periods that minimise the dynamical heights and wind velocities. 
Rotation periods of 16.58 h for Uranus and  17.46 h for Neptune were derived. 
 Although there is no underlying principle to strictly enforce dynamical height minimization, the principle provides very good results when applied to Jupiter and Saturn. For Jupiter, it leads to a rotation period in agreement with that of its system III (Helled et al. 2009), and in Saturn, the rotation period inferred is in excellent agreement with the ones inferred from atmospheric vorticity arguments (Read et al. 2009), ring seismology (Mankovich et al. 2019) and the measured gravitational field (Helled et al. 2015, Militzer et al. 2019). 
 The different rotation periods of Uranus and Neptune as inferred from Helled et al.~(2010) can lead to significant differences in the inferred composition and internal structure of the planets, as shown by Nettelmann et al. (2013). 
In addition, it was shown by Kaspi et al. (2013), based on a thermal wind model and interior models, that in both planets the observed winds are not expected to penetrate very deep: for a penetration depth  above $\sim 1000$\,km, the influence on $J_4$ would have been 
larger and incompatible with any of the structure models that fit $J_2$. \\

\noindent{\bf Physical Shapes}\\
The mean radius and flattening are used as constraints for interior models. While knowledge of the continuous shape of a planet (i.e., radius vs. latitude) is also desirable and available, typically, structure models use only the flattening ($f =(R_{eq}-R_p)/R_{eq}$ where $R_{eq}$ and $R_p$ are the equatorial and polar radii, respectively), which is derived for given polar and equatorial radii. 
The shapes of Uranus and Neptune have been studied through stellar and ring occultations, and Voyager measurements. Stellar and ring occultations (e.g., French et al., 1983, 1987, 1998, Hubbard et al., 1987, Elliot \& Nicholson, 1984) provide very good determinations of the planetary shape, however, the data are limited to low pressure-levels  from several to 10$^3$ mbar, and therefore do not necessarily apply to the typical 1-bar pressure-level that is used by interior modelers, in particular, because atmosphere dynamics can change the isobaric shape between the microbars and 1 bar pressure-levels. Therefore an uncertainty of up to $\sim$100 km in radius should be considered when modeling the interiors of these planets.  
Clearly, a determination of the continuous planetary shape can be used to further constrain structure models. 
Figure 1 shows the inferred shapes of Uranus and Neptune assuming the Voyager 2 and modified solid-body rotation periods for the planets as derived by Helled et al.~(2010). \\
 It is important to note that the atmospheric winds (and dynamics in general) can significantly modify the planetary shape, and this effect can be larger than the harmonic coefficients. As a result, knowledge of the planetary shape has only a limited effect on internal structure models. Nevertheless, the atmospheric shape is important for understanding the dynamics of the planetary atmospheres.\\

\begin{figure}
  \includegraphics[angle=0,height=4.12cm]{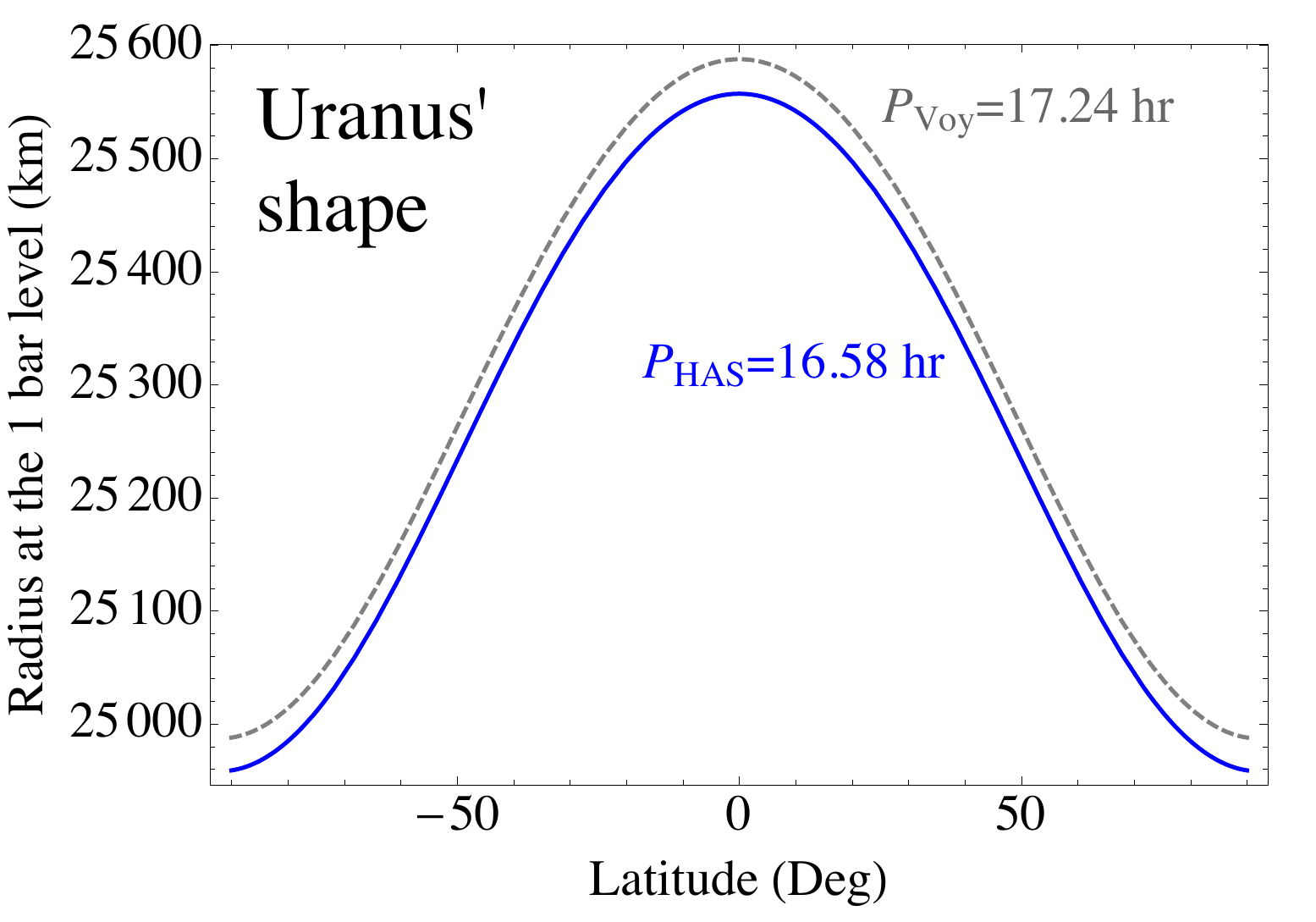}
    \includegraphics[angle=0,height=4.1cm]{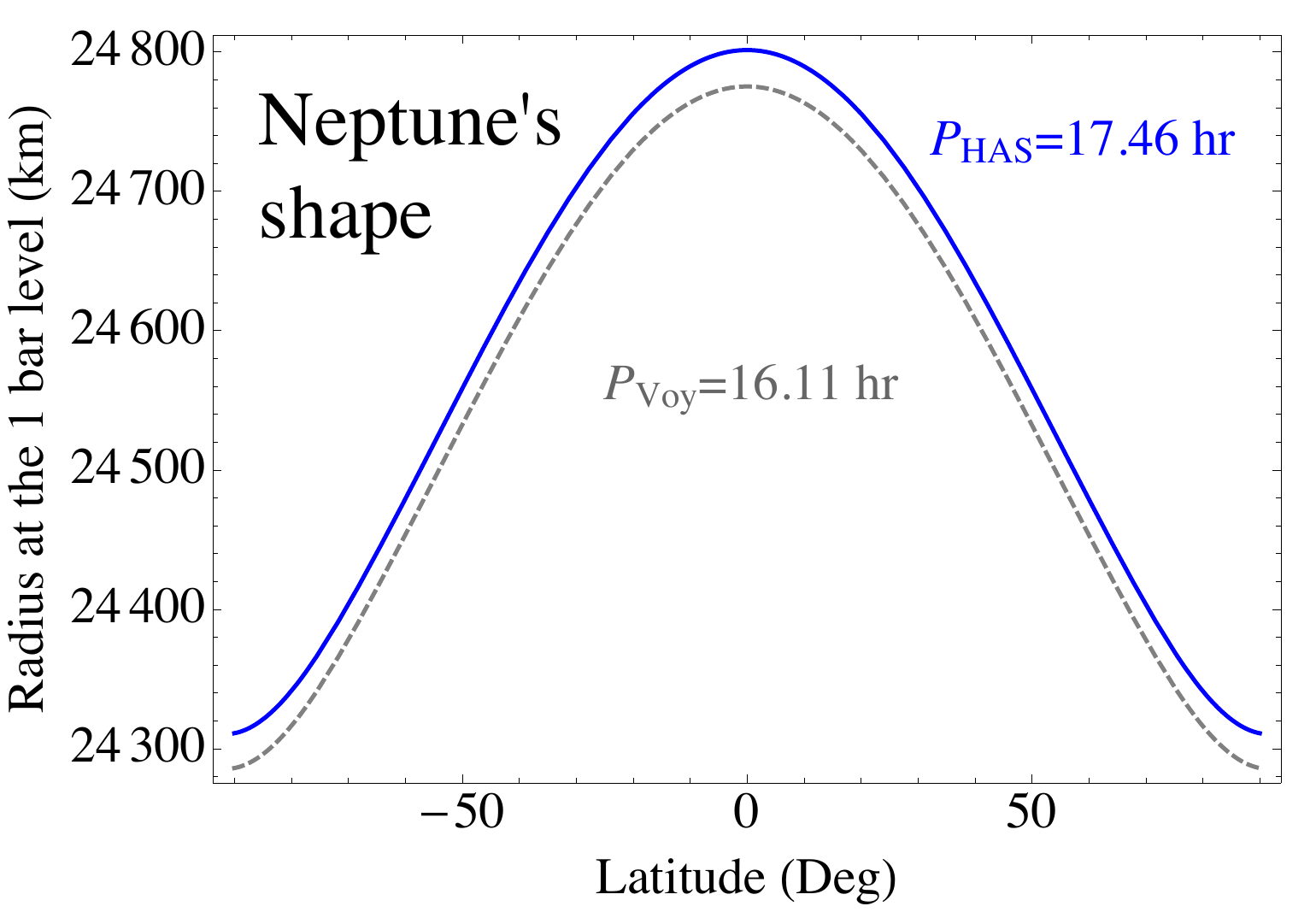}
\caption{The physical shapes of Uranus and Neptune. The dashed-gray and curves correspond to the  Voyager rotation periods (U: 17.24 hrs, N: 16.11 hrs) while the solid blue are the shape corresponding to rotation periods that minimizes the planets' winds and dynamical heights as found by Helled at al. (2011)  (U: 16.58 hrs, N: 17.46 hrs). }
\label{fig:1}       
\end{figure}

\noindent{\bf Atmospheres}\\
Uranus is typically seen as a planet with a bland, featureless atmosphere while Neptune is seen as more active, with vortexes comparable to those seen on Jupiter. This suggests that Uranus, which has a ten times lower heat flux is less convectively active than Neptune. This view must however be taken with caution because it is largely based on the high-resolution but short visits from the Voyager 2 spacecraft in 1985 and 1989, respectively 
(Allison et al., 1991, West et al., 1991, Ingersoll et al., 1995, Baines et al., 1995). Ground based observations with large telescopes have since shown that both planets have dynamic atmospheres, with features appearing and disappearing regularly (e.g., Karkoschka2011, Sromovsky et al., 2012, dePater et al., 2015, Molter et al. 2019). Unfortunately, the lack of a high spatial resolution achievable from an Earth-based observatory combined to the limited amount of time over which these facilities can be pointed to these targets strongly limits our ability to understand these planets globally.

The visible atmospheres of Uranus and Neptune are dominated by three species: hydrogen, helium and methane. While the helium to hydrogen ratio seems consistent with a protosolar abundance, methane is very abundant, reaching a volume mixing ratio of around 2\% corresponding to an enrichment of 50 to 100 over the solar value (Guillot \& Gautier, 2014 and references therein).  Methane condenses at pressures below 1.5 bar in both planets (Lindal, 1992), and probably form the haze and some of the clouds that are observed. Most other key species condense at higher pressures (and temperatures) and are thus mostly hidden from sight. Radio-observations capable of probing the deep atmospheres of Uranus and Neptune have provided some constraints on the presence of H$_2$S, and a lack of NH$_3$ (dePater et al., 1991). This lack of NH$_3$ has been confirmed indirectly by the detection of H$_2$S in the atmosphere or Uranus (Irwin et al., 2018) and with a high likelihood, Neptune (Irwin et al., 2019).  
 Water cannot be in solution in the gas phase until pressures of at least 90 bars, far from direct detection. 
 Recently, using ALMA, Tollefson et al. (2019) were able to show that both H$_2$S and CH$_4$ are not distributed uniformly across latitudes on Neptune, which has implications to understand the atmospheric dynamics of the planet.
\par

The temperature profiles derived from Voyager 2 radio occultations (Lindal, 1992) are characterized by a minimum (tropopause) near 0.1 bar with 53.0\,K for Uranus and 51.7\,K for Neptune. The temperature is increasing with depth (higher pressures) to reach 76.4\,K and 71.5\,K at 1\,bar, and deeper 101\,K at 2.3\,bar  and 135\,K at 6.3\,bar for Uranus and Neptune, respectively. This increase is consistent with a dry adiabat but the degeneracy between temperature and methane abundance implies that many solutions are possible, with possible super- or sub-adiabatic gradients (Guillot, 1995). The uncertainty on the assumed temperature profile (or equivalently, deep entropy) is significant and adds to the uncertainty on possible interior structures and compositions (see Leconte et al., 2017, Friedson \& Gonzales, 2017). 

Unfortunately, water is mostly in condensed form and therefore in extremely small abundances until pressures of at least 90 bars, too deep to be detectable. Constraints on the water abundance can be obtained indirectly, from the detection of CO in the atmosphere: CO is out of equilibrium and must be supplied from deeper region where the reaction $H_2O + CH_4  \xleftrightarrow{} CO + 3H_2$ can take place. The constraint depends on a diffusion coefficient to account for the transport of CO and on a temperature profile. Using an eddy diffusion coefficient from mixing length theory and superadiabatic temperature profiles obtained from Leconte et al. (2017), Cavail{\'e} et al. (2017) determine that, nominally, O/H should be $<$ 160 times solar in Uranus and about 540 times solar in Neptune. Varying the diffusion coefficient and the assumed temperature profile results in considerable variations in those values. The fact that they differ significantly between the two planets is puzzling: it may be due to real composition differences or to differences in the mixing - since Uranus is indeed much less active than Neptune this seems to be a more plausible explanation. More data are needed to understand the deep atmospheric structures of these planets in order to make progress on this important issue. 
More information on the atmospheric composition of Uranus and Neptune is inferred from measurements of their deuterium to hydrogen D/H ratios. 
The measurement D/H from observations by the Herschel spacecraft in the far infrared yields a value that is similar for both planets, i.e. $(4.4\pm 0.5)\times 10^{-5}$ for Uranus and $(4.1\pm 0.1)\times 10^{-5}$ for Neptune (Feuchtgruber et al., 2013). This value is intermediate between the one found in Jupiter (2.25$\pm$0.35)$\times 10^{-5}$ (e.g., Lellouch et al. 2001) and values obtained from Oort cloud comets, i.e., $2-3\times 10^{-4}$ (Bockel{\'e}e-Morvan et al., 2012). 
Assuming the planets were once fully mixed, the rather low atmospheric D/H enrichment led Feuchtgruber et al. (2013) to propose, based on interior models from Helled et al.~(2011) and Nettelmann et al.~(2011), that Uranus and Neptune 
should be made of 68\% to 86\% of rocks instead of being mostly icy. It should be noted, however, that it is rather unlikely that the planets were once fully mixed (see below for further details). 
Further studies of comets have shown that the D/H ratio can be as low as $1.5\times 10^{-4}$ (Lis et al., 2019), the relatively low observed D/H ratio it likely to be the outcome of partial mixing of the volatile materials in the planets. 
There is a clear need for more atmospheric measurements and for a better knowledge of how to connect the atmosphere but clearly such measurements can be used as an additional constraint for future interior models. 
\\

\noindent{\bf Magnetic fields}\\
A key observable property is the planetary magnetic field. 
Structure models must be consistent with the observed multi-polar magnetic fields, implying that 
a convective and electrically conductive region of a width of $\sim 20\%$ of the planetary radius exist underneath the outer H-He-rich envelope (e.g., Stanley \& Bloxham, 2004, 2006, Redmer et al.,2011).  
The latter is insulating and transitions into the observable atmosphere. 
Dynamo models that fit the Voyager magnetic field data suggest that the deep interior below the dynamo region is stably stratified (Stanley \& Bloxham 2004, 2006) or, alternatively, in a state of thermal-buoyancy driven turbulent convection (Soderlund et al., 2013). 
Since Voyager's observations have not been confirmed by another spacecraft, it is unclear whether the Voyager rotation rate reflects the rotation of the layer in which the magnetic field is generated and of the entire deep interior below that region (Helled et al., 2010). This uncertainty has major consequences on the inferred planetary structure and the question of similar or dissimilar interiors (Nettelmann et al., 2013). 
Improved measurements of the magnetic fields of Uranus and Neptune will also help to constrain the planetary rotation rate and internal structure. 
The available physical parameters of the planets are summarized in Table 1. 

\begin{table}[h!]
\def\arraystretch{1.5}
\centering 
\begin{tabular}{lll}
\hline
\hline
{\bf Parameter} & {\bf Uranus} & {\bf Neptune}\\
\hline
Mass ($10^{24}$ kg) & 86.8127 $\pm$ 0.0040$^a$& 102.4126 $\pm$ 0.0048$^b$\\
Mean Radius$^*$ (km) & 25362 $\pm$ 7$^c$ & 24622 $\pm$ 19$^c$\\
Mean Density (g cm$^{-3}$) & 1.270 $\pm$ 0.001$^d$& 1.638 $\pm$ 0.004$^d$\\
R$_{\rm ref}$ (km) & 26,200$^a$& 25,225$^b$ \\
J$_2$ ($\times$10$^{6}$) &3510.68 $\pm$ 0.70$^a$& 3408.43 $\pm$ 4.50$^b$\\
J$_4$ ($\times$10$^{6}$) & -34.17 $\pm$ 1.30$^a$ & -33.40 $\pm$ 2.90$^b$\\
$P_{\rm Voy}$ (rotation period) & 17.24h$^e$& 16.11h$^f$\\
$R_{\rm eq,Voy}$ (km)  & 25,559 $\pm$ 4$^c$ & 24,764 $\pm$ 15$^c$ \\
$R_{\rm p,Voy}$ (km) & 24,973 $\pm$ 20$^g$ & 24,341 $\pm$ 30 $^g$ \\
$P_{\rm HAS}$ (rotation period) & 16.58h$^g$& 17.46h$^g$\\
$R_{\rm eq,HAS}$ (km)  & 25,559 $\pm$ 4$^g$ &24,787 $\pm$ 4$^g$  \\
$R_{\rm p,HAS}$ (km) & 25,023 $\pm$ 4$^g$ & 24,383 $\pm$ 4$^g$  \\
1-bar Temperature (K) & 76 $\pm$ 2$^h$& 72 $\pm$ 2$^h$\\
Effective Temperature (K) & 59.1 $\pm$ 0.3$^i$ & 59.3 $\pm$ 0.8$^i$ \\ 
Intrinsic flux (J s$^{-1}$ m$^{-2}$) & 0.042$^i$ $\pm$ 0.045 & 0.433 $\pm$ 0.046$^i$\\
Bond Albedo $A$ & 0.30 $\pm$ 0.049$^i$ & 0.29 $\pm$ 0.067$^i$ \\
Magnetic dipole moment (Tm$^3$) &  3.9$\times10^{17}$$^j$  &  2.2$\times10^{17}$$^j$ \\
\hline
\hline
\end{tabular}
\caption{Physical data of Uranus \& Neptune. 
$^a$Jacobson, R.A. 2014. $^b$Jacobson, R.A. 2009.
$^c$Archinal et al. 2018. $^d$Calculated values and associated uncertainty derived from other referenced values and uncertainties in this table. The average density is computed using a  volume of a sphere with the listed mean radius. 
$^e$Desch et al., 1986 $^f$Warwick et al., 1989. $^g$Helled et al., 2010. $^h$Lindal, 1992. 
$^i$Pearl \& Conrath, 1991. $^j$Russell \& Dougherty, 2010.
$^*$Note that the listed uncertainties in mean radius are the formal measured ones, and they do not account for the uncertainty in shape and rotation periods, which leads to a higher uncertainty as discussed in the text. 
R$_{ref}$ is the reference equatorial radius in respect to the measured gravitational harmonics J$_2$ and J$_4$, $R_{eq}$ and $R_p$ are the equatorial and polar radii at the 1-bar pressure level, respectively.}
\label{tab:1}       
\end{table}

\section{The Compositions and Internal Structures of Uranus \& Neptune}

\subsection{Constraints from structure models}
The composition of the planets cannot be measured directly (besides in the very upper atmosphere) but has to be inferred indirectly from interior models fitting all available constraints.  
Structure models are designed to fit the measured physical parameters of the planets (mass, radius, gravitational field, rotation rate, atmospheric temperature), and have a density profile that reproduces the measured gravitational coefficients. 
Traditional interior models of Uranus and Neptune assume an ``adiabatic'' structure all the way to the interior, i.e., a temperature profile set by the specific entropy of hydrogen and helium measured in the atmosphere (typically at 1\,bar).  The idea is that, as for Jupiter, convection should dominate and the super-adiabaticity required to transport the internal heat flux is small. It is important to realize however that in the presence of compositional gradients (or boundary layers), this definition is no longer correct. It requires heat transport to proceed efficiently (so that the temperature is continuous across composition interfaces) while chemical elements are not transported at all, an unlikely situation. Departures from the ``adiabatic'' hypothesis may explain the differences in densities and luminosities of Uranus and Neptune: their inner temperature may be higher than traditional models predict, and their luminosity could be controlled by the location and characteristics of the regions with changing composition (e.g., Podolak et al. 1995, Vazan \& Helled, 2019). The thermal evolution and profiles of the planets are discussed in detail in Section 6.

A first approach uses physical equations of state (EoSs) of the assumed materials to derive the density and the associated pressure and temperature (Hubbard et al., 1991; Podolak et al., 1995, Nettelmann et al., 2013). The second approach uses empirical (mathematical) density profiles without making a priori assumptions regarding the planetary structure and composition and are not linked to specific EoSs and yet fits all the measurements (Marley et al. 1995, Podolak et al. 2000, Helled et al., 2011). 
 These "empirical" models provide a density-pressure profile of the planet. This can then be interpreted using physical EoS for different materials, and assuming various temperature profiles, in order to infer the planetary composition.   

Adiabatic  interior models of Uranus and Neptune that are based on physical EoSs, in which the rocks are confined to the core predict small core 
masses and large ice mass fractions, leading to highly super-solar ice:rock ratios of at least $4\times$ 
the solar value for Neptune (Nettelmann et al., 2013) and about $15\times$ the solar value for Uranus (e.g., Podolak \& Reynolds, 1987; Nettelmann et al., 2013), suggesting that some additional fraction of rocky materials are also 
mixed into the icy envelope. 

\begin{figure*}
\center
  \includegraphics[angle=0,height=7.1cm]{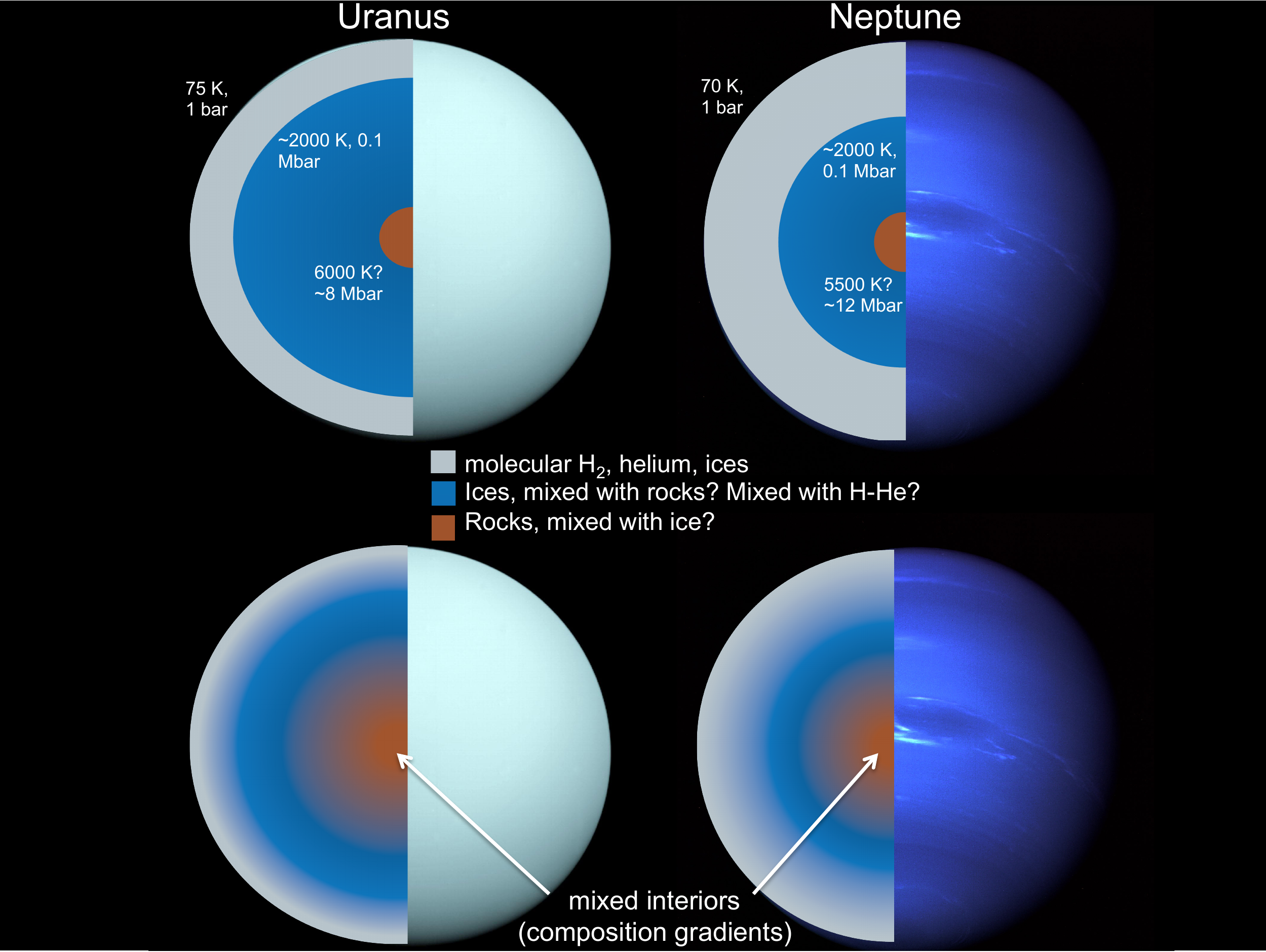} 
\caption{Possible structure model for Uranus and Neptune based on Helled et al., 2011, Nettelmann et al., 2013.}
\label{fig:2}       
\end{figure*}

In the mid-90s it was found that in order to fit Uranus' gravity field, the density in the ice shell must be 10\% lower than the one given by the EoSs used at that time (Podolak et al.,1995). In addition, the inferred ice-to-rock ratio in this model was 30 by mass, roughly 10 times the proto-solar ratio.  
Recent interior models for Uranus (Bethkenhagen et al., 2017) using the updated gravity data 
(Table 1) confirm the earlier obtained possibility (Podolak \& Reynolds 1987)
of a nearly pure-ice shell. The high resulting ice-to-rock ratio of the deep interior for this class of models can be reduced if there is a super-adiabatic transition between between the  H/He-rich outer and heavy-element
rich interior. The deep interior then becomes significantly hotter than in the adiabatic case, and the water is in a plasma phase where the EoS is sensitive to temperature, and the increase in volume must be compensated for by a significant amount of rocks, allowing even for solar ice-to rocky ratios (Nettelmann et al., 2016).

\begin{figure*}[h!]
\center
  \includegraphics[angle=0,height=7.5cm]{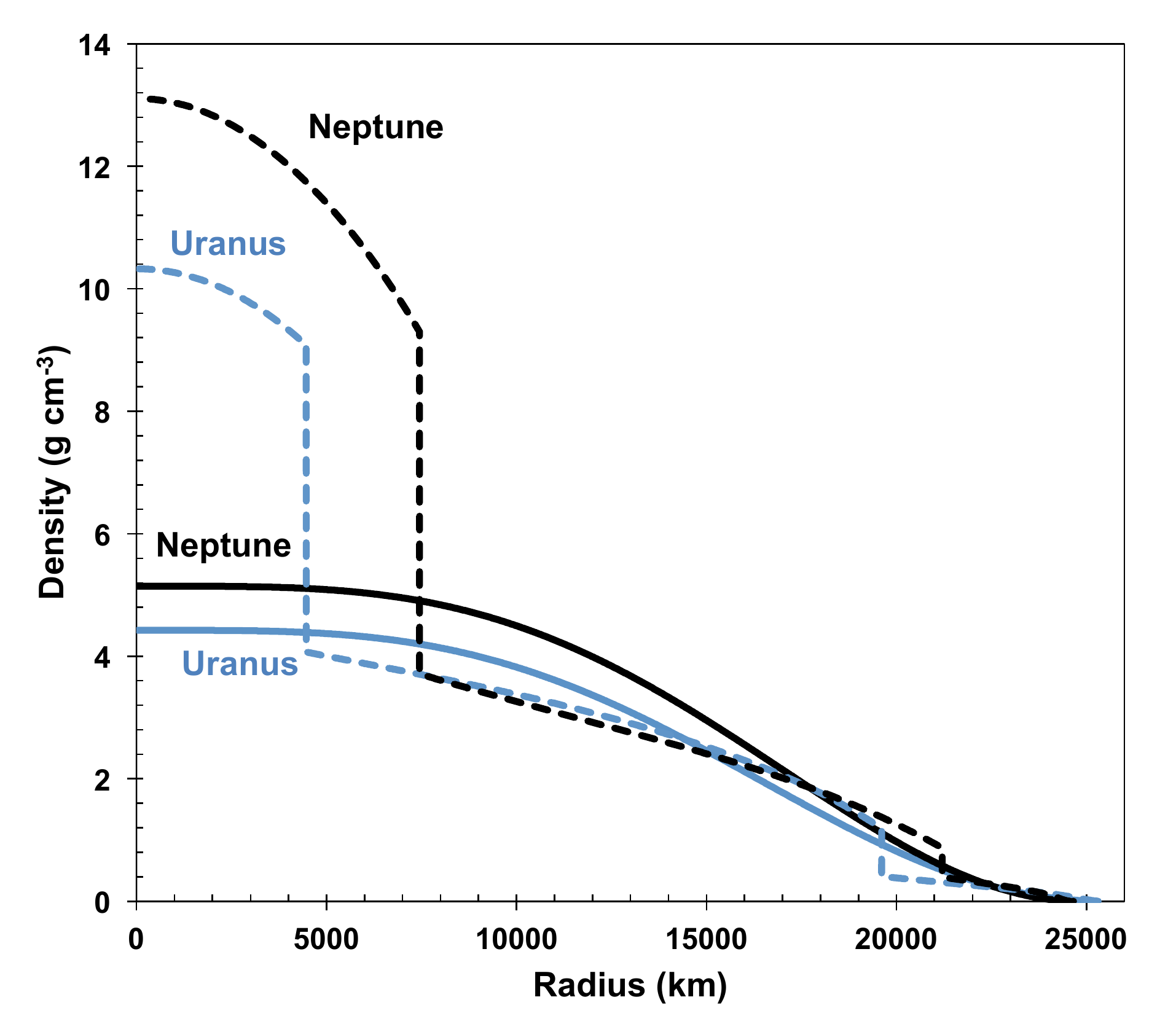}
\caption{Density as a function of radius for Neptune (black) and Uranus (blue). The solid curves are the density profiles presented in Helled et al. (2011). The dashed curves are for the three-layer models of Nettelmann et al. (2013).}
\label{fig:2}       
\end{figure*}

A range of density profiles of Uranus and Neptune that fit their measured gravitational fields were derived using Monte Carlo searches by Marley et al.~(1995) and Podolak et al. ~(2000). These random models imply that both planets consist of small cores and outer envelopes enriched with heavy elements, and that both planets consist of a density discontinuity at a radius of $\sim$0.6-0.7. 
Helled et al. (2011) represented the density profile ($\rho(r)$) of Uranus and Neptune by a 6th-order polynomial, and have found the coefficients required to fit all of the observed properties. It was shown that a density profile with non-distinct layering (i.e., a density profile without
discontinuities) can also satisfy the observational constraints. 
The "empirical" EoS generated by these models was then interpreted as requiring a continuous increase in the H-He mass fraction towards the planetary center. 
It was shown that the gravity data can be fit as well with silicates as with water. 
When comparing the inferred density-pressure profile from these models to physical EoSs, it was shown that the planets do not need to contain large fractions of water to fit their
observed properties (e.g., Helled et al., 2011). It was then concluded that the interior structures of Uranus and Neptune are poorly understood, that they could be rock-dominated, and that their interiors may differ from the "standard 3-layer models". Figure 2 shows sketches of the possible internal structure solutions for the planets, and Fig. 3 shows possible density profiles for the planets as inferred by these studies.  

Water-rich 3-layer adiabatic models predict a metallicity of $\sim 85-87\%$ for Uranus and of $81-84\%$ for Neptune (Nettelmann et al., 2013). 
Interior models that are based on empirical density profiles, suggest a metallicity of  $\sim 76\%$  to $\sim 90\%$ for Uranus and $77\%$ to $90\%$ for Neptune, when the heavy elements  are  being represented by SiO$_2$ and water (Helled et al., 2011).  
Non-adiabatic interior models for Uranus and Neptune were recently presented by Podolak et al.~(2019) who explored how the assumption of non-adiabatic temperature profiles in the planets affects their internal structures and compositions.  Various plausible temperature profiles were used together with density profiles that match the measured gravitational fields to derive the planetary compositions. 

It was found that the inferred compositions of both Uranus and Neptune are quite sensitive to the assumed thermal profile in the outer layers, but relatively insensitive to the thermal profile in the central, high-pressure region. The heavy-element mass fraction for both planets was found to be between 0.8 and 0.9, in agreement with other structure models of the planets (e.g., Helled et al., 2011; Nettelmann et al., 2013). 
 This result is linked to the behaviour of hydrogen as hydrogen gas has a very large adiabatic lapse rate due to its low molecular weight. As a result, even a very small (in mass) H-He atmosphere can imply high interior temperatures, if an adiabatic temperature profile is assumed. 
{\it The inferred global ice-to-rock-ratio in Uranus and Neptune is in fact unknown, and depends on the model assumptions and the materials that are chosen to represent the heavy elements}. In addition, Uranus and Neptune could have complex deep interiors that are dominated by composition gradients and/or phase boundaries. At the moment, it is fair to say that our understanding of the compositions and internal structures of the planets is incomplete. 
These may be a result of demixing in the cool, mature planet (e.g., Wilson \& Militzer, 20111) and/or from the formation process (e.g., Helled \& Stevenson, 2017).

A better understanding of the interior could arrive from EoS calculations and phase diagrams. Internal structure models must be consistent with the phase diagram of the assumed materials and their mixtures. 
Detailed information on EoS calculations and the connection to planetary interiors can be found in Helled et al.~(2018) and references therein. 
Calculating the EoS of different materials for the conditions existing in the interiors of Uranus and Neptune is very challenging because molecules, atoms, ions and electrons coexist and interact, and the pressure and temperature cover several orders of magnitude, with the pressures going up to several mega-bars (Mbar), i.e., 100 GPa and the temperature can reach up to 10$^4$ Kelvins. 
As a result, a deep understanding of the planetary interiors requires to perform high-pressure experiments and solve the many-body quantum mechanical problem of the system. Therefore there is a clear connection between modeling planetary interiors and high-pressure physics. 
For example, it is possible that Uranus and Neptune have deep water oceans that begin where H$_2$ and H$_2$O become insoluble  (e.g., Bailey \& Stevenson, 2015, Bai et al., 2013)  or that some of the materials become miscibile in planetary conditions  (e.g., Soubiran \& Militzer, 2015).   
Figure 4 presents sketches of four possible internal structures of the ice giants where the transitions between layers distinct (via phase/thermal boundary) and/or gradual.

\begin{figure*}
\center
  \includegraphics[angle=0,height=3.5cm]{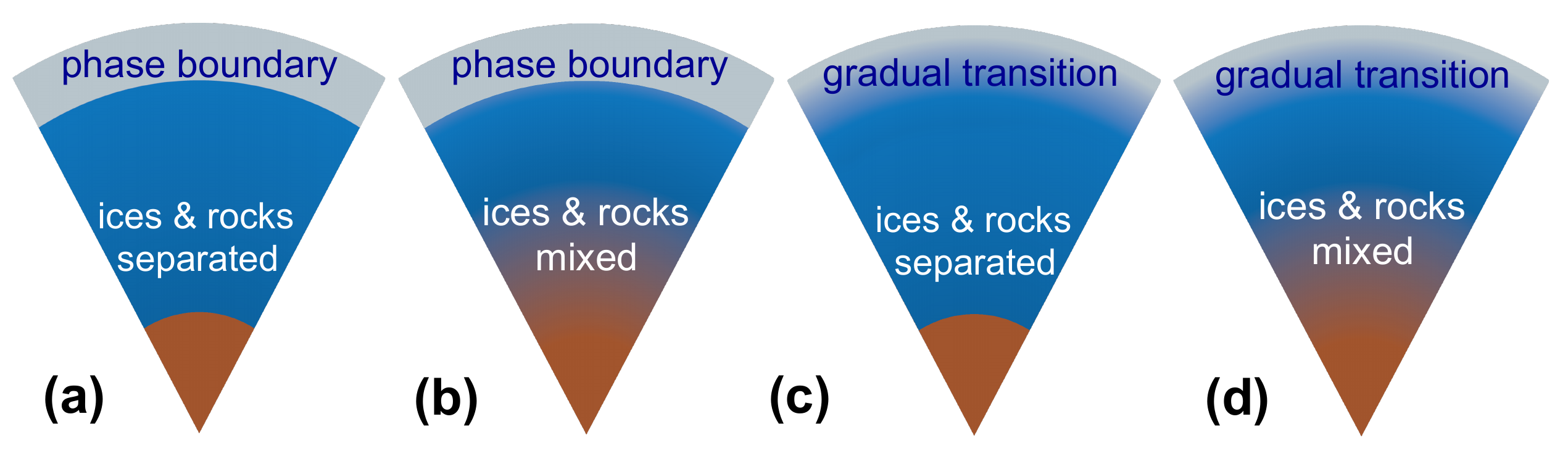}
\caption{Sketches of the possible internal structures of an ice giant. It is unclear whether Uranus and Neptune are differentiated and whether the transition between the different layers are distinct or gradual: (a) separation between the ices and rocks and the ice and H-He atmosphere (b) separation (phase boundary) between the H-He atmosphere and ices and a gradual transition between ice and rock, (c) gradual transition between the H-He atmosphere and ice layer, and a distinct separation between the ice and rock layers, and (d) gradual transition both between the H-He atmosphere and ice and the ice and rocks suggesting a global composition gradient with the planets (see text for discussion). }
\label{fig:2}       
\end{figure*}

\subsection{Are Uranus \& Neptune really "icy" planets?}
Uranus and Neptune are often referred as the "ice giants". 
While some internal structure models predict that the planets are highly enriched with water, as discussed above it is not a unique solution, and in fact, there are reasons to question whether they are truly "icy" worlds. 
There are several arguments as to why the planets are expected to be water-rich:\\
(1)  Uranus and Neptune are located at large radial distances of about 20 and 30 AU, where the temperatures in the solar nebula are expected to be low enough to create water-rich solids (pebbles/planetesimals) that are accreted by the planets;
(2) Oxygen is very abundant in our Sun and the ice (water and volatile materials condensing at temperatures of order 100 to 300\,K) to rock ratio is expected to be between 2 and 3 (Lodders 2003). 
(3) Uranus and Neptune have magnetic fields, implying the existence of conductive material, which was suggested to be ionic water (e.g., Nellis et al., 1997, Redmer et al., 2011). 

However, these arguments may be challenged. As discussed above, all the observed parameters can be reproduced  if the innermost regions of the planets consist of  $\sim$82\% rock by mass with the rest being a mixture of hydrogen and helium in proto-solar ratio (Helled et al., 2011). Of course this is an extreme case, which is rather implausible, as water is also expected to be present, but it clearly demonstrates that the available data do not directly imply that Uranus and Neptune have high water-to-rock ratios. Also, we cannot exclude the existence of other materials that can result in high enough conductivity to generate a magnetic field such as  compressed silicates, especially when mixed with hydrogen (e.g., Soubiran \& Militzer, 2018). 
In addition, the oxygen-to-hydrogen ratios in the atmospheres of Uranus and Neptune are not well-determined, although despite the large uncertainties, 
the existing measurements do indicate high oxygen-to-hydrogen ratios in their atmospheres, as inferred from the observed CO abundance,
ranging between a few and a few hundred times the proto-solar ratio (Luszcz-Cook \& dePater, 2013). 
Finally, we cannot exclude that formation models yield an ice to rock ratio in the protosolar disk that differs from that in the Sun (e.g., Ida \& Guillot, 2016). After all, Pluto, which is located even farther out in the solar system, contains about 70\% rocks (e.g., McKinnon et al., 2017).

\subsection{The connection between atmosphere and interior}

A proper characterization of the atmosphere is crucial for modeling the planetary interior structure and evolution and constrain the planet's composition. 
Because giant planets are fluid and have no obvious surfaces (except perhaps very deep down), atmospheric and interior composition are intimately linked. The thermal structure of the atmosphere, including possible latitudinal variations, impacts directly modeling of the deep interior and constraints on bulk composition and core mass that may be derived. Last but not least, the atmosphere is the lid governing the planet cooling (e.g., Guillot, 2005).

One major unknown concerning the atmospheres of Uranus and Neptune is the way their internal heat flux is transported. The large molecular weight of condensible species compared to the background gas, hydrogen and helium, implies that moist convection can be inhibited past an abundance of the condensing species over about 6 to 10\% in mass (Guillot, 1995). This critical abundance is in fact reached by methane in both Uranus and Neptune, and by water, if the C/O ratio is less than about four times solar. Furthermore, the criterion is not affected by double-diffusion (Leconte et al., 2017, Friedson \& Gonzales, 2017). This implies that there is considerable uncertainty on the inner temperature profile for both planets as illustrated for the water-condensation region in Fig.~5.\\
 At the moment, we are still at a stage that there is no match between the observed atmospheric abundances of the planets and the ones derived with chemical models (e.g., Cavalie et al. 2017) with the rather low values predicted from structure models. As discussed above, structure models suffer from degeneracies, and in terms of atmospheric composition, it is still unclear whether the atmospheric composition represents the planetary bulk. 

\begin{figure}[h!]
\center
\includegraphics[width=0.75\textwidth]{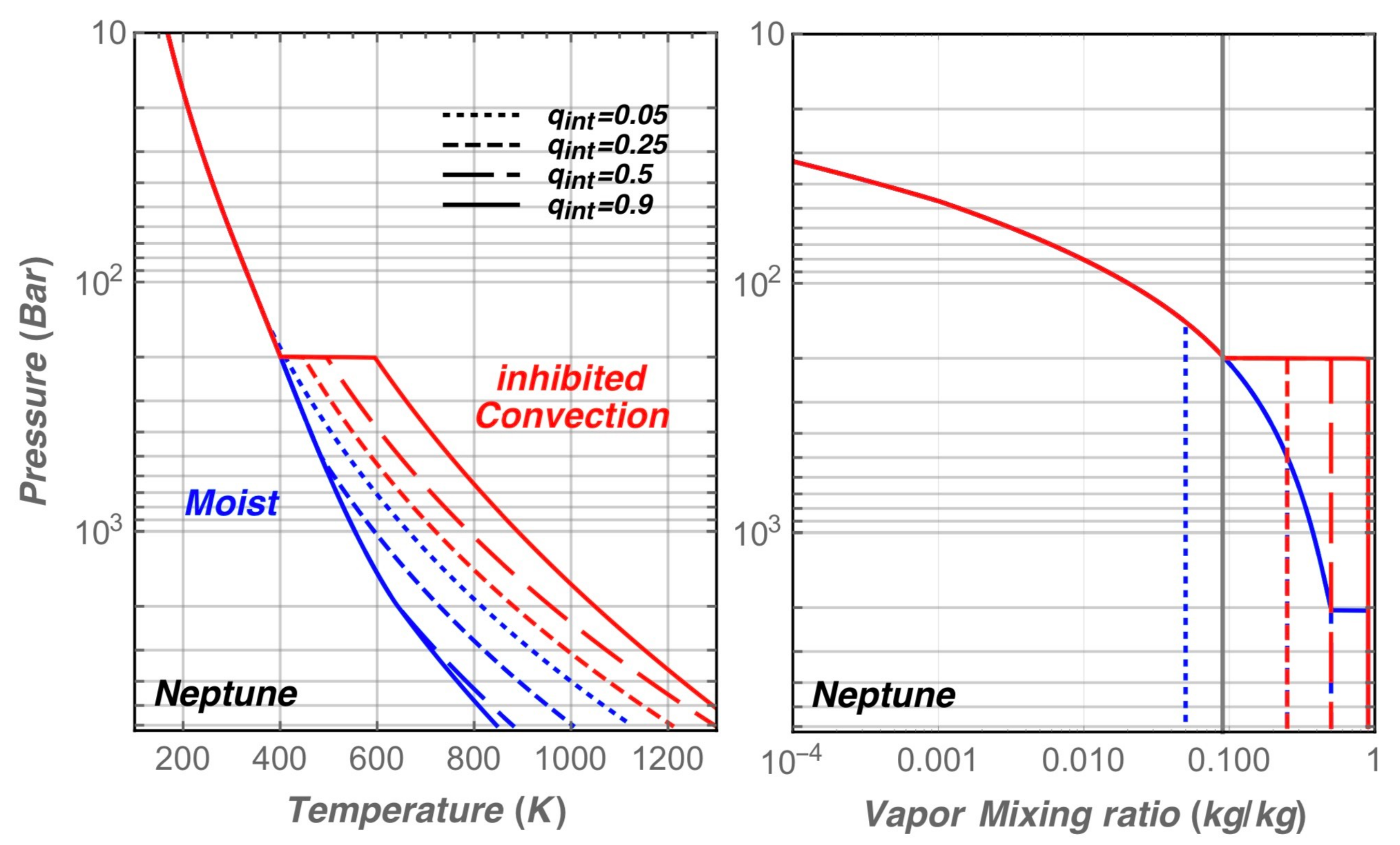}
\caption{Possible temperature and water abundance profiles in the region of water condensation in Neptune. {\bf Left:} Temperature vs.~pressure. {\bf Right:} Water vapor abundance vs.~pressure. The different lines correspond to different assumed bulk mass mixing ratios of H$_2$O, from 0.05 to 0.9. Blue curves correspond to the usually assumed moist convective profile. The red curves correspond to a situation in which heat transport is done by radiation where moist convection is inhibited. Figure from Leconte et al., 2017.}
\label{fig:leconte2017}
\end{figure}

\section{The Formation of Uranus \& Neptune}
\label{sec:2}
The formation of Uranus and Neptune has been a long-standing problem for planet formation theory (e.g., Pollack et al., 1996, Dodson-Robinson \& Bodenheimer, 2010, Helled \& Bodenheimer, 2014, Frelikh \& Murray-Clay, 2017). 
At the same time, the large number of detected exoplanets with sizes comparable (or smaller) to that of Uranus and Neptune suggests that such planetary objects are very common, at least at short orbital distances (e.g., Batalha et al. 2013, Petigura et al., 2018). 

In the standard planet formation model, core accretion (see Helled et al., 2014 for review and the references therein), a slow planetary growth is expected to occur at large radial distances where the solid surface density is lower, and the accretion rate of planetesimals is significantly smaller (e.g., Helled \& Bodenheimer, 2014). For the current locations of Uranus and Neptune, the formation timescale can be comparable to the lifetimes of protoplanetary disks.
In addition, forming the planets {\it in situ} requires extremely high solid surface densities, which has led to the idea that the planets formed closer to the sun and reached their current locations at later stages (e.g., Thommes et al. 1999). 
Due to the long accretion times at large radial
distances, the formation process is too slow to reach rapid gas accretion (runaway), before the gas disk disappears, leaving behind an intermediate-mass planet, which
consists mostly of heavy elements and a small fraction of H-He gas. 
However, since the mass of H-He in both Uranus and Neptune inferred from structure models is estimated to be a couple of M$_{\oplus}$, 
it implies that gas accretion has already begun, and this requires that the gas disk disappears at a very specific time, to prevent further gas accretion onto the planets. 
This is known as the {\it fine-tuning} problem in Uranus/Neptune formation.

Helled \& Bodenheimer (2014) investigated the formation of Uranus and Neptune in the core accretion model accounting for different formation locations ranging from 12 to 30 AU, and with various disk solid-surface densities and core accretion rates. 
This systematic study confirmed that in order to form Uranus and Neptune with the correct final planetary mass and solid-to-gas ratio, very specific conditions ({\it fine-tuning}) are needed. 
It was also shown that the potential high-accretion rates associated with pebble accretion (e.g., Lambrechts \& Johansen, 2012) and dynamically cold planetesimal disks (e.g., Rafikov, 2011) at large radial distances (several AU) can result in much shorter formation timescales for the planets. 
 At the same time, a recent N-body calculation by Levison et al.~(2016) that accounts for viscously stirred pebble accretion suggests that Uranus and Neptune formed at shorter radial distances (5-15 AU). It is still unclear what are the favourable formation locations for Uranus and Neptune and at the moment it is not possible to discriminate among the different models. 
 In any case, dynamical models focus on the growth rate of the heavy-elements and the main challenge  in formation models of Uranus and Neptune is to reproduce the final mass and composition of the planets, accounting for the accretion of both the solids and gas, and find mechanisms that prevent the planets from becoming gas giants. 
The challenge in forming Uranus and Neptune is demonstrated in Fig.~6 which shows various formation scenarios that lead to the formation of planets similar to Uranus and Neptune in terms of mass and composition. 
Shown are the planetary mass the H-He mass and the mass of heavy elements.

\begin{figure*}
\center
  \includegraphics[angle=0,height=8.5cm]{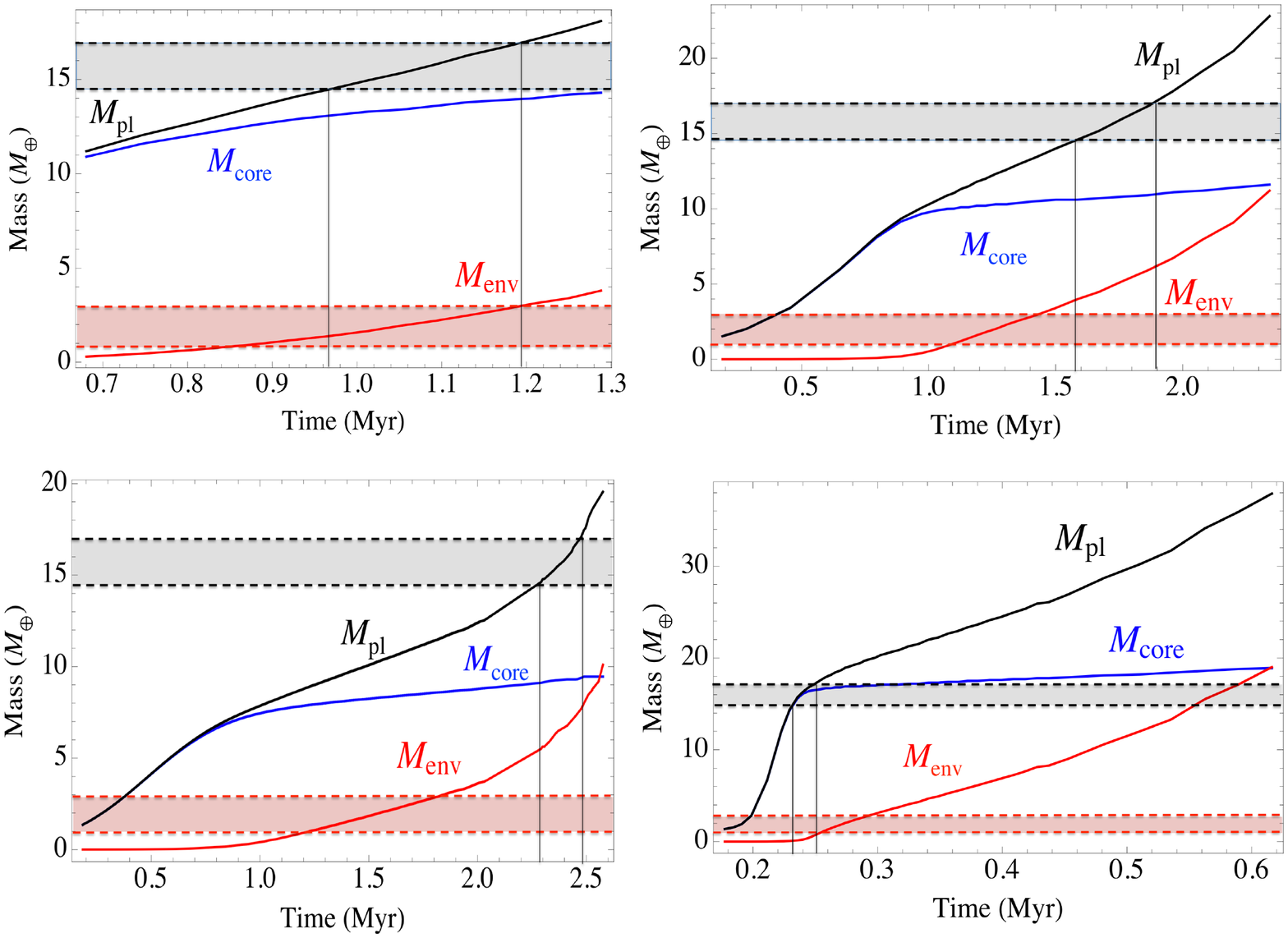}
\caption{Formation paths of Uranus and Neptune. 
The dashed black curves correspond to the masses of Uranus and Neptune, and the red dashed llines to the inner and upper bound of the H-He mass. The vertical black lines show when the planet reaches Uranus/Neptune mass and the $M_{core}$ (heavy-element mass, blue curve) and $M_{env}$ (hydrogen-helium mass $M_{H-He}$), and the total planetary mass (black curve $M_{pl}$) in that time. Acceptable models are ones in which the black horizontal lines are within the gray and red area. Figure modified from Helled \& Bodenheimer, 2014.}
\label{fig:2}       
\end{figure*}

As can be seen from the figure there are several challenges:
First, the forming planets should not become gas giants, second, the planets should have accreted some H-He gas, but not in amounts that exceed the upper bounds inferred from structure models, i.e., the planetary metallicity should be of the order of 80-90\%.
It can be seen that even for these preferred models, which assume high accretion rates and/or smaller radial distances for the planets 
as suggested by the Nice model (e.g., Thommes et al. 1999; Tsiganis et al. 2005), it is hard to reproduce the right masses and $M_{H-He}$ to $M_Z$. 
On the other hand, the study of Helled \& Bodenheimer (2014) demonstrated how small changes in the properties of the protoplanetary disks and the birth environment of the planetary embryos can lead to the formation of very different planets in terms of final masses and compositions (solid-to-gas ratios), which naturally explains the large diversity of intermediate-mass exoplanets. 
\newline
\indent
Recently it was shown that when enrichment of the H-He envelope with heavy elements is included, gas accretion is expected to take place faster making the formation of Neptunes even more challenging  (Venturini et al. 2016, Venturini \& Helled, 2017). 
A mechanism that prevents rapid gas accretion onto intermediate-mass protoplanets is required to explain the formation of Uranus and Neptune as well as Neptune-like planets and mini-Neptunes (e.g., Alibert et al., 2018). 
Another possible formation path as suggested by Lambrechts et al.~(2014) is that Uranus and Neptune grew by pebble accretion. In this case the planets can form {\it in situ} within a few Mys. This is because in that scenario, the core growth is more efficient than in the planetesimal accretion case, and at the same time, at the current locations of the planets the pebble isolation mass is above M$_{\oplus}$. As a result, the planets could be heavy-element dominated with H-He envelopes that are metal-rich due to the sublimation of icy pebbles (see Lambrechts et al., 2014 for further details). While forming Uranus and Neptune via pebble accretion might seem appealing, 
it should be noted that until the growing planet reaches the pebble-isolation mass the H-He mass fraction is {\it assumed} to be 10\% of the total mass (e.g., Bitsch et al., 2015). This assumed 10\% H-He mass fraction of the pebble accretion model  was found to be unrealistically low, suggesting that this formation scenario would lead to too high masses of H-He (Venturini \& Helled, 2017). Currently, the pebble accretion model has not yet shown a self-consistent formation path for these planets. Finally, another formation path for Uranus and Neptune is formation by collision and merging of a few low-mass planets which accreted from a population of planetary embryos, which significantly decreases their formation timescale (e.g., Izidoro et al. 2015).  Each model seems to have weaknesses and strengths, and yet, there is not satisfactory formation model for Uranus and Neptune. 
\newline
\indent
Measuring the elemental abundances in the atmospheres of Uranus and Neptune can provide information on the formation history of the planets, by setting limits on their formation locations and/or the type of solids (pebbles/planetesimals) that were accreted by the planets as discussed above.  
For example, it was shown by Kurosaki \& Ikoma (2017) that the pollution of the protoplanetary atmospheres with heavy elements (in particular water, ammonia and methane) can significantly affect the cooling of the growing planet and therefore its formation history as well as final internal structure.  
In addition, a determination of the atmospheric metallicity will provide valuable constraints for structure models, that at the moment allow a large variation of this value as described above. 
Therefore, in order to understand the formation of these planets direct measurements of their atmospheric composition are required.

\section{Giant impacts on Uranus and Neptune}
\label{sec:impacts}

Uranus and Neptune are similar in terms of masses and radii but also have significant differences such as their heat fluxes, their satellite systems, and possibly also their internal structures.  
Giant impacts by large embryos occurring shortly after the formation of the planets could explain the dichotomy between the ice giants (Safronov 1966, Stevenson 1986). 
An oblique impact with a massive impactor could change Uranus' spin (Safronov 1966), and at the same time eject enough material that will result in the forming of a disk where the regular satellites could form. 
An oblique impact is expected to mostly affect the angular momentum of the planet, but not its internal structure. Therefore, if Uranus was differentiated and/or  consisting of boundary layers, such an event is unlikely to affect the deep interior. 
For Neptune, it is suggested that the collision was head-on, which could reach the planetary deep interior, "erasing" the distinct layers, possibly also eroding the core, and lead to a "more convective \& mixed" interior, which is consistent with its measured heat flux and inferred MoI value. 
Figure 7 presents a sketch of this scenario.  \\

\begin{figure}[h!]
\center
\includegraphics[width=0.67\textwidth]{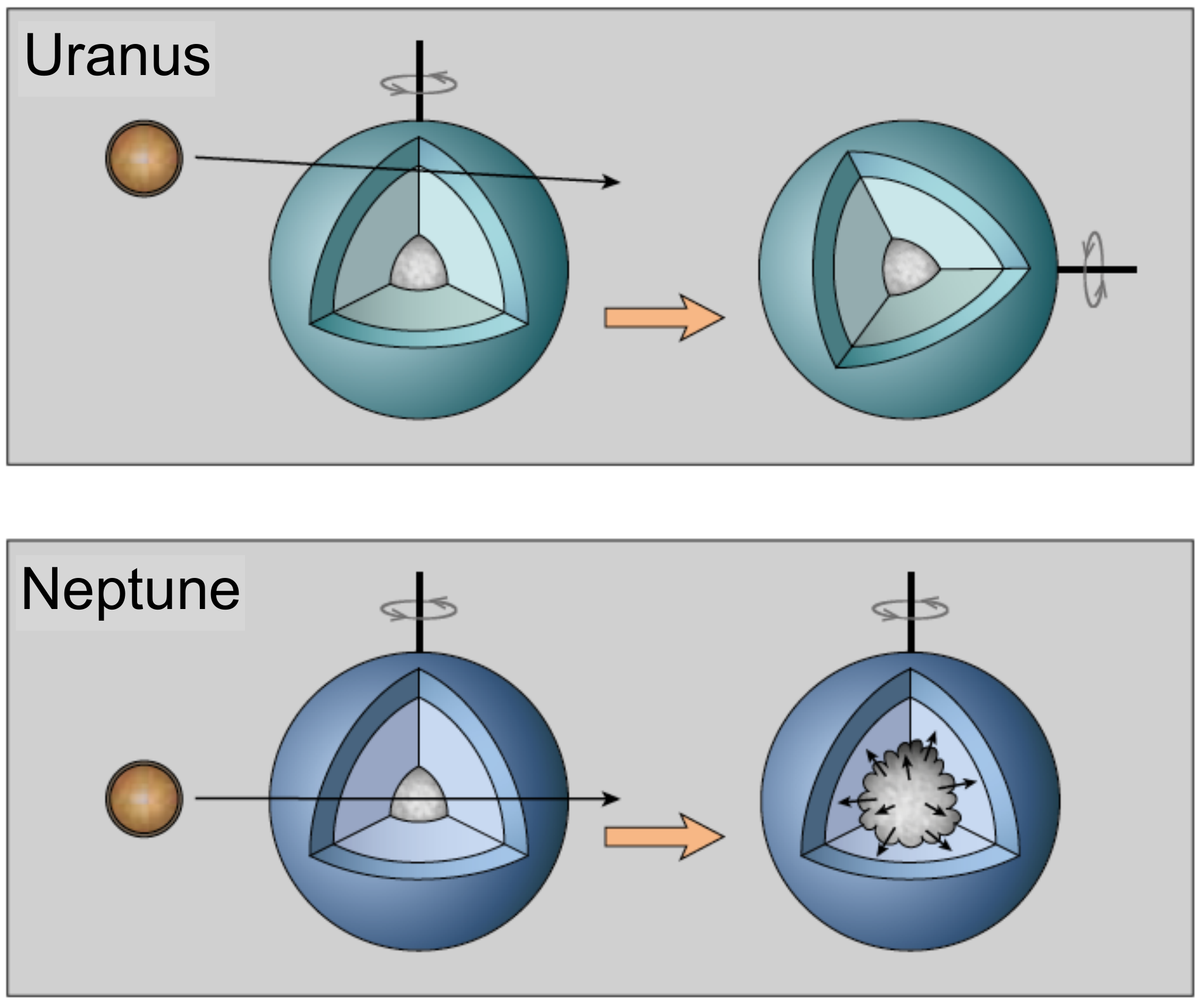}
\caption{\small 
A sketch presenting the idea of the role of giant impacts in explaining the dichotomy between Uranus and Neptune (not to scale). 
An oblique giant impact on Uranus could tilt its spin axis significantly and eject enough material to form a disk and the regular satellites while keeping the body stratified. 
For Neptune, an almost head-on collision might deposit energy deep inside, mixing its interior resulting in a thermal profile that is close to adiabatic explaining the fast cooling. From Jaumann et al.~(2018), adapted from R.~Helled, based on Podolak \& Helled~(2012).}
\end{figure}


Podolak \& Helled, (2012) performed a simple analysis of this scenario, and estimated the energy and angular momentum exchange of large impactors, 
 and showed that head-on collisions, which add relatively little angular momentum to the planet can have sufficient energy to mix large fractions of the core, while oblique collisions can add large amounts of angular momentum without affecting the core. 
These results are in agreement with the original idea proposed by Stevenson (1986). 
\par

Recently several studies investigating giant impacts on Uranus and Neptune using Smoothed Particles Hydrodynamics (SPH)  have been presented (Kegerreis et al., 2018, 2019, Kurosaki \& Inutsuka, 2018, Reinhardt et al. 2019). 
A large parameter space (impact geometry, impactor's mass and composition, and numerical parameters) 
was considered to identify the collisions that can reproduce the observed properties of Neptune and Uranus (Reinhardt et al., 2019). 
Studies of Uranus confirmed that an oblique impact 
can alter its rotation period, tilt the spin axis, and eject enough material to create a disk where the  regular satellites are formed. 
For Neptune, it was confirmed that massive and dense projectiles can penetrate towards the center and affect its interior. 
This could 
lead to an 
adiabatic temperature profile, which explains its larger flux and higher moment of inertia value. 
For both planets the rotation axes and periods can be reproduced based on these simulations. 

\begin{figure}[h!]
\center
\includegraphics[width=10cm, height=7cm ]{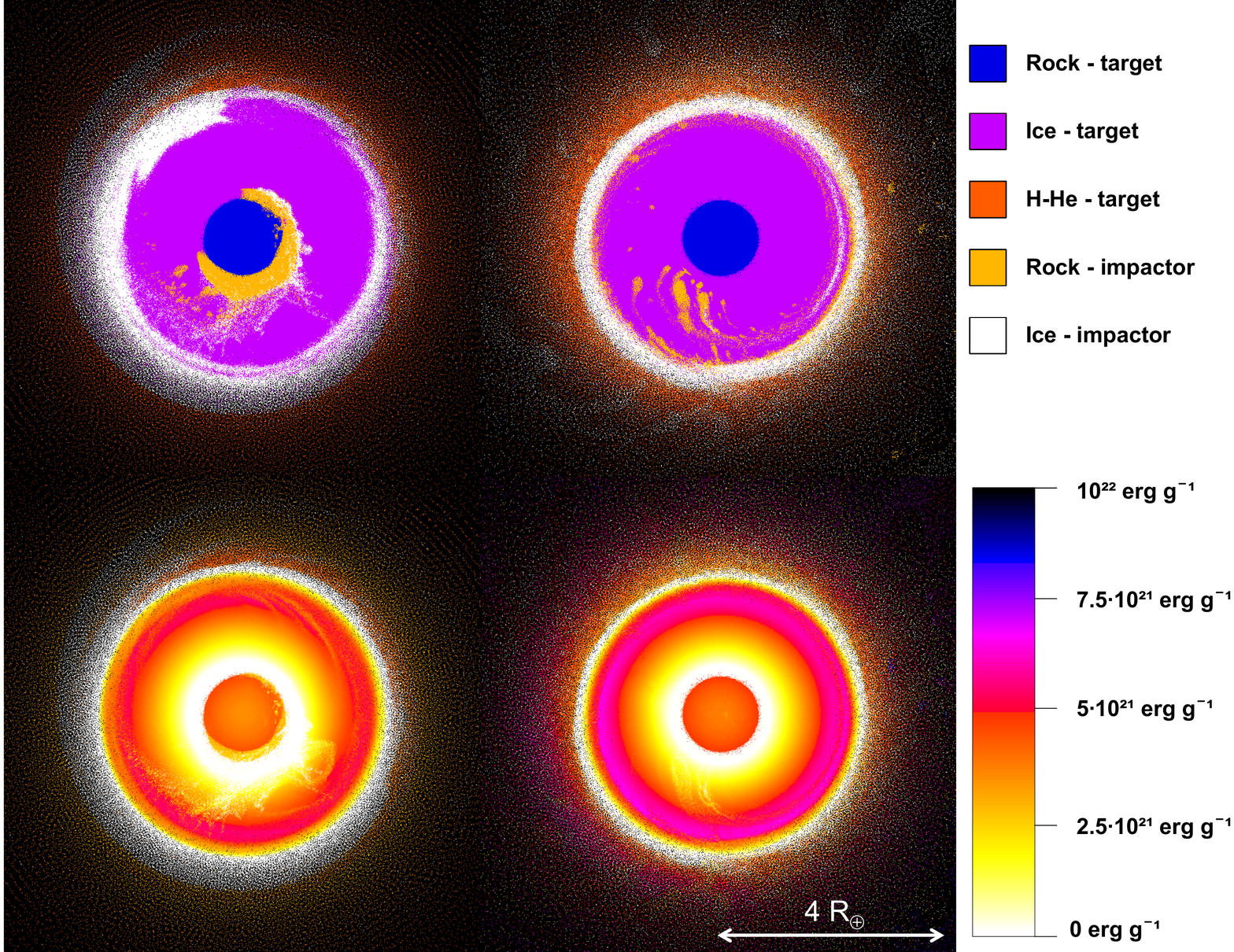}
\caption{\small 
The planetary interior after a head-on ($b=$0.2, left) and a grazing ($b=$0.7 , right) collisions on an ice giant. Shown are the results for a differentiated 
2$M_{\oplus}$ projectile colliding with proto-Uranus using $5 \times 10^6$ particles at $v_{\inf}$=5km/s 71h after the impact. 
The panels shows the origin of the material (top) 	
	and the specific internal energy (bottom) as indicated by the colorbar. 
	For the head-on collision the impactor penetrates into the planetary deep interior, and the planet is substantially heated. 
	In case of the grazing collision the impactor interacts with the planet's outer regions, survives the first encounter (not shown) 
	and is substantially tidally eroded before the second impact. 
	Therefore much less material and energy are deposited in the planet. 
	Figure adapted from Reinhardt et al., 2020.}
\end{figure}

Figure 8 compares the outcome of a head-on collision and a grazing collision on a Uranus-like planet consisting of a small rocky core, a water envelope, and a H-He atmosphere as presented by Reinhardt et al.~(2020).  Shown are the materials and internal energy. While a head-on collision affects the internal structure by depositing mass and energy in the deep interior, a grazing collision does not significantly  affect the internal structure. 
The pre- and post- angular momentum distributions are listed in Table 2.

\begin{table}[h]
\begin{center}
\def\arraystretch{1.5}
\begin{tabular} {||c|c|c|c|c|c||}
        \hline
        Impact parameter & $L_{initial} [erg\cdot s]$ & $L_{final} [erg\cdot s]$ & $L_{final} [erg\cdot s]$ & $L_{initial}/L_{final}$ & $L_{initial}/L_{final}$ \\ 
        $b$  & target+impactor & planet+envelope & planet & planet+envelope & planet \\
        \hline
        0.2 & 1.295$\times 10^{43}$ & 1.116$\times 10^{43}$ & 5.087$\times 10^{42}$ & 0.86 & 0.39 \\ 
        \hline
        0.7  & 4.535$\times 10^{43}$ & 1.613 $\times 10^{43}$ & 7.567 $\times 10^{42}$ & 0.36 & 0.17 \\  \hline
    \end{tabular} 
    \caption{Angular momentum $L$ for the head-on ($b=0.2$) and grazing ($b=0.7$)  impacts shown in Figure 8. Listed are the initial ($L_{initial}$) and final ($L_{final}$) angular momenta 
    for the planets and the surrounding   material ("envelope"). 
    The angular momentum is evaluated at the centre of mass which does not always correspond to the planet's center. }
    \label{tab:truthTables}   
    \end{center}
\end{table}

While further investigations of the topic and modeling the long-term thermal evolution of the post-impact planets are  required, 
the following conclusions can already be made: 
(i)  Giant impacts can explain some of the observed differences between Uranus and Neptune.
(ii) Giant impacts on Uranus and Neptune can substantially alter their rotation axis and internal structure.
(iii) A giant impact on Uranus can lead to the formation of an extended disk providing enough material for the formation of its regular satellites after the collision. 
(iv) Head-on collisions for Neptune result in accretion of more mass and energy, and substantially affect the planetary interior.

These studies represent only the beginning of a long-term exploration of the role of giant impacts in understanding Uranus and Neptune.
While these studies are encouraging, they do not prove that the observed differences between the planets are indeed caused by giant impacts, and alternative explanations are still possible such as orbital migration (e.g., Bou\'{e} \& Laskar 2010). 
Nevertheless, the recent giant impact studies strengthen the idea that giant impacts  play an important role in determining the planetary properties not only in the inner part of the Solar System (Mercury, Earth's moon) but also in the outer part. 

Finally, it is interesting to note that giant impacts on Uranus and Neptune could also assist formation models.
For example, if Neptune suffered a head-on collision leading to an accretion of most of the impactor's mass, it would naturally increase Neptune's mass by the impactor's mass, for which massed between 1-3 $M_{\oplus}$ have been considered, and therefore would also increase the $M_Z$ to $M_{H-He}$ ratio. 
Indeed, the Nice model suggests the formation of two $\sim$15 $M_{\oplus}$ planets at radial distances of 12 and 15 AU, where Uranus and Neptune are indistinguishable (e.g., Thommes et al. 1999; Tsiganis et al. 2005).

\section{Long-term Thermal Evolution of Uranus and Neptune}
\label{sec:2}

In this section we address the thermal evolution of Uranus and Neptune, shortly after their formation until today. 
The duration it takes to cool from the luminous, hot, and extended initial state to their observed luminosity 
and radius is denoted the cooling time $\tau$, and should be consistent with the age of the solar system.  

Under the assumption of an adiabatic interior after the run-away phase of core accretion (see section 
3), Marley et al.~(2007) showed that the memory of the initial state is lost after only 10 Myrs for a 
1~$M_{\rm Jup}$ H-He planet, and that this timescale decreases the lower the mass of the planet.  
However, this study corresponded to gaseous planets which are homogeneously mixed and adiabatic, and therefore might not apply for Uranus and Neptune. 
In a non-adiabatic interior, as expected for the ice giants, the memory of the initial state including giant impacts can be preserved over Gyr timescales.   
The fact that adiabatic models loose rapidly the memory of their initial conditions whereas this is not necessarily the case for non-adiabatic ones can be understood as follows: 
For adiabatic models, the temperature gradient within the planet follows the adiabatic gradient  and is entirely defined by its atmospheric boundary condition. 
Therefore the contraction and cooling  of the entire planet goes approximately as $1/T_{eff}^3$ (Hubbard 1977) with the exact relation depending on the 
 atmospheric opacities. Therefore the cooling time (i.e., the long-term evolution timescale) for the adiabatic case does not depend on the initial condition: 
 primordial hotter interiors cool faster than colder ones and reach a similar internal state relatively fast. This is not the case for a non-adiabatic structure where the temperature gradient differs from the adiabatic one. 
 In non-adiabatic models, the temperature profile in different parts of the interior can differ significantly  from the adiabatic one and are decoupled from the atmosphere through the presence of radiative, conductive or semi-conductive zones (see Guillot et al. 1995, Leconte \& Chabrier 2012, Vazan et al. 2016). As we discuss below, composition gradients and boundary layers within the ice giants result in planetary cooling timescales significantly longer than that planet's current age and the planet is still relaxing from the initial condition. 
Cooling times of adiabatic models of Uranus and Neptune are 
reviewed in section \ref{sec:evolUN_ad}, models with stable stratification are discussed in Section 
\ref{sec:evolUN_stable}, and models with a non-adiabatic deep interior are discussed in Section 6.4.


\subsection{Luminosity and Effective Temperature}
The cooling of fluid planets is largely governed by their composition in terms of H-He, ices (water, ammonia, methane), and refractory materials (metals, silicates), by the internal distribution of these components, and by external and internal heat sources. 
Cooling times of {\it adiabatic models} of Uranus and Neptune are found to be rather insensitive to the composition distribution as long as the mass fractions are chosen to reproduce the measured gravity field (Fortney et al., 2011, Nettelmann et al., 2013), or even only their mass and radius (Hubbard, 1978; Linder et al., 2019).
Adiabatic evolution models have therefore been computed assuming a structure with silicates and iron confined
to the core, a middle layer of ices, and an outer H-He envelope (Hubbard \& MacFarlane, 1980, Fortney et al., 2011; Linder et al., 2019),  
or with a middle layer of ices enriched in H-He and an outer H-He-layer enriched in ices (e.g., Hubbard et al., 1995, Fortney \& Nettelmann, 2010, Nettelmann et al., 2013).

The long-term evolution of (weakly irradiated) planets can be calculated by integrating over time the energy 
balance equation 
\begin{equation}
	L_{\rm eff} - L_{\rm eq} = L_{\rm int} = L_{\rm sec} + L_{\rm radio},
\end{equation}
where $L_{\rm eff}=4\pi R_p^2\,\sigma_B\,T_{\rm eff}^4$ is the observable luminosity and 
$L_{\rm int}=4\pi R_p^2\,\sigma_{\rm B}\,T_{\rm int}^4$ is the heat loss from the interior. Its major
contribution results from  cooling and contraction described by 
$L_{\rm sec}=-4\pi R_p^2\,\int_{\Mcore}^M \!dm\, T\,\Delta s/\Delta t + L_{\rm core}$, where $dm\,T(m) \Delta s$ 
is the heat lost by the envelope mass shell $dm$ at $m$. Other luminosity sources could be added, but the ones mentioned are the dominating ones in the case of Uranus and Neptune. 

As discussed in Section 3, if the planets have rock-dominated interiors, 
radiogenic heating from the rocky component, $L_{\rm radio}$, can prolong the cooling 
time of adiabatic Neptune by up to 0.4 Gyr while for adiabatic Uranus by a several Gyrs (Nettelmann et al., 2016).
The core contribution $L_{\rm core}$ is rather small and therefore it matters little whether it is assumed to be adiabatic or isothermal (e.g., Linder et al., 2019).  Finally, $L_{\rm eq}=4\pi R_p^2\,\sigma_{\rm B}\,T_{\rm eq}^4$ is the absorbed and re-emitted irradiation. 
Its value depends on the Bond albedo $A$, which is estimated from \textit{Voyager} (see Table 1). As with the rock mass fraction, its value is of weaker influence on $\tau_{\rm Nep}$ while its $1\sigma$ uncertainty changes 
$\tau_{\rm Ura}$ by a significant amount of $\pm 0.5$ Gyrs (Scheibe et al., 2019). 

The generally stronger response of $\tau_{\rm Ura}$ than of $\tau_{\rm Nep}$ in adiabatic models is because 
the effective temperature $T_{\rm eff}=59.1\pm 0.3\:$K of Uranus is close to its equilibrium value 
$T_{\rm eq}=58.1\pm 1.1\:$K. Since $T_{\rm int}^4 = T_{\rm eff}^4 - T_{\rm eq}^4$ and $L_{\rm int} 
\sim T_{\rm int}^4$, the observations based finding of $T_{\rm eff}\simeq T_{\rm eq}$ for Uranus implies 
that little to no heat escapes from the present planet ($L_{\rm int}\sim0 $). Small changes in external (Albedo) or internal 
(radiogenic) heat sources therefore have a relatively large effect on $T_{\rm int}$ and Uranus' cooling time.

\subsection{Adiabatic models}\label{sec:evolUN_ad}

The adiabatic assumption is probably inappropriate for modeling the evolution and internal structures of Uranus and Neptune. 
Nevertheless, such simple models can reveal the uncertainties in our knowledge and guide the development of more complex models. 
In adiabatic models, the specific entropy $s$ is assumed to be constant within layers of homogeneous 
composition. Between the layers, the entropy may change as a result of the change in composition. 

\begin{figure}[h]
\centering
\includegraphics[width=9cm]{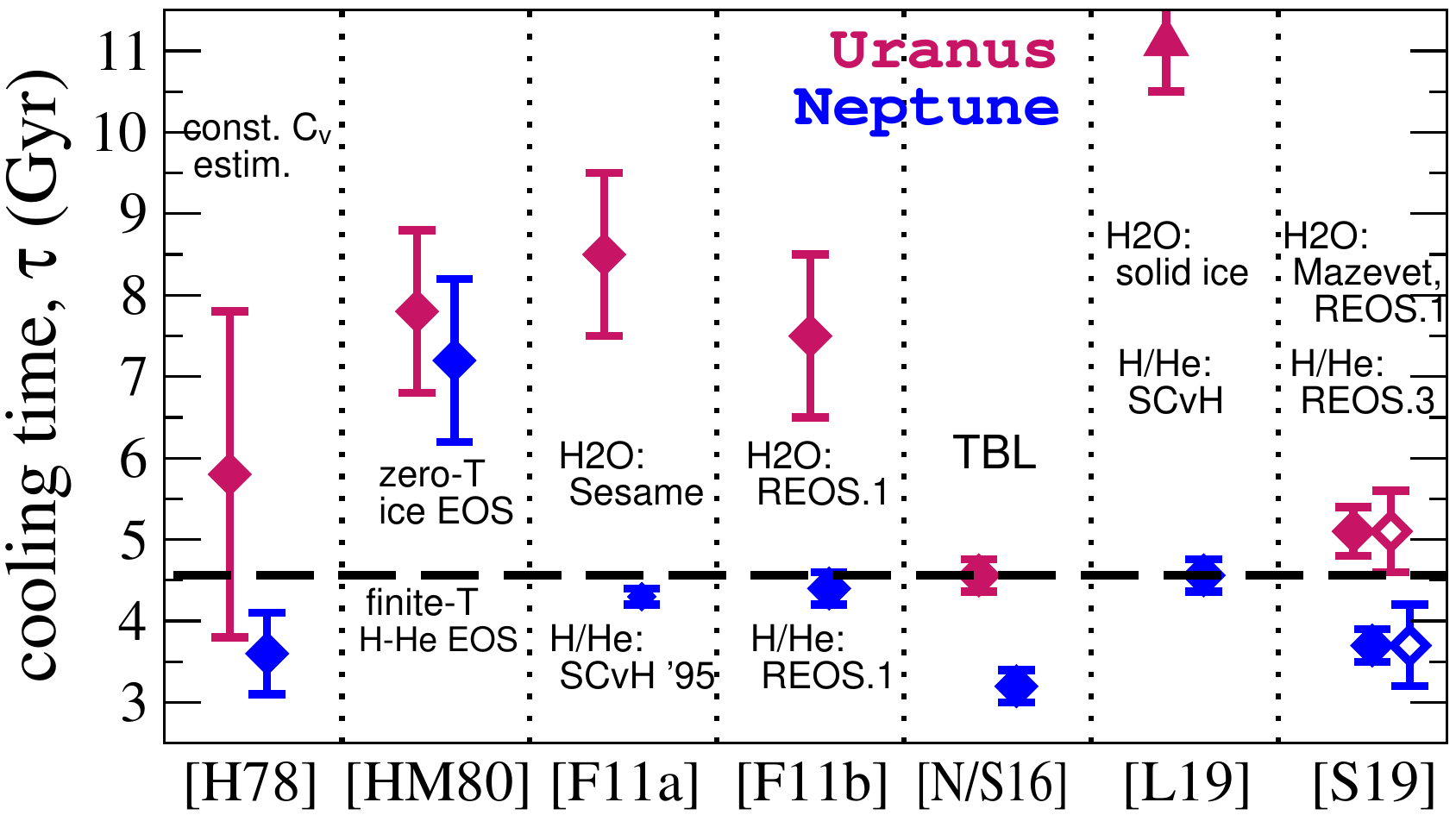}
\caption{\label{figNN:taus}Cooling times to reach the present luminosity of Uranus (red) and Neptune (blue). 
assuming an adiabatic interior except for the model labeled TBL. Refs.:
[H78]: Hubbard (1978) assuming Jupiter model scaled in radius, specific heat $C_v$, and Gruneisen
$\gamma$, 
[HMf80]: Hubbard \& MacFarlane (1980) using zero-T EOS fitted to exp.~data for different ices, 
[F11a]: Fortney et al.~(2011) using Sesame water EOS, fully differentiated layers, constant $T_{\rm eq}$,
[F11b]: using H2O-REOS and mixed layers, constant $T_{\rm eq}$, 
[N/S16]: Nettelmann et al.~(2016) for Uranus and L.~Scheibe (Master Thesis 2016, U Rostock) using $T_{\rm eq}(t)$ 
and assuming super-adiabatic TBL,
[L19]: Linder et al.~(2019) using solid state EOSs and constant $T_{\rm eq}$, 
[S19]: Scheibe et al.~(2019) using $T_{\rm eq}(t)$.
}
\end{figure}

Adiabatic models of Uranus take longer to cool to the present luminosity than adiabatic models of Neptune 
and longer than the age of the Solar system of $\tau_{\odot}=4.56\:$Gyr. This general finding is illustrated 
in Figure \ref{figNN:taus}. How much $\tau_{\rm Ura}$ exceeds $\tau_{\odot}$ depends on the equations of state 
used and on the assumed temporal behavior of $T_{\rm eq}$, which is often assumed to be constant although 
its value increases over time due to the evolution of the Sun. The property $T_{\rm int}\simeq 0$ of Uranus together with $\tau_{\rm U}\gg\tau_{\odot}$ is known as the faintness problem of Uranus (e.g., Podolak et al., 1991).   

Cooling times of an adiabatic Neptune have been found longer, equal, or shorter than the age of the solar system. 
In particular, results obtained over the past decade agree about $\tau_{\rm Nep} \leq \tau_{\odot}$, see
Figure \ref{figNN:taus}. 
By fine-tuning the ice-to-rock ratio and the Albedo within its $2\sigma$ uncertainty it is possible to 
find evolution models that yield $\tau_{\rm Nep}=\tau_{\odot}$ for a wide range of considered H-He and 
water equations of state, suggesting that Neptune's interior is largely convective and adiabatic.
At present it remains an open question whether models with $\tau_{\rm Nep} < \tau_{\odot}$ indicate an excess 
luminosity of Neptune, or a rock-rich interior and low Bond Albedo.

\subsection{Models with a thermal boundary layer}\label{sec:evolUN_stable}

Uranus' luminosity  can be brought into agreement with the age of the solar system if one
assumes that some fraction of the interior is shut off from efficient cooling due to stable stratification
(Podolak et al., 1991, Hubbard et al., 1995), in which case a thermal boundary layer (TBL) would develop at the transition to
the convectively cooling, adiabatic outer region. 
However, the location and behavior of a thermal boundary layer inside Uranus, if there is any, is not well-known. 
Nettelmann et al.~(2016) {\it assumed} that the TBL occurs due to a composition gradient between the H-He envelope
and the "icy" interior at about 80\% of its radius, and that the super-adiabatic temperature
gradient across grows monotonously over time. This yields the solution labeled [N/S16] in Figure \ref{figNN:taus}
for Uranus.

Within the wide range of structure models that are possible for Neptune due to the larger uncertainty in the measured 
gravitational harmonics (see Table 1), similar internal structures of Uranus and Neptune are not excluded\footnote{although the likely different giant impact histories suggest dissimilar
interiors even if they shared the same formation history (see section 4).}. 
Moreover, the most different (adiabatic) internal structures, which may be 
considered as a sign of dichotomy (Nettelmann et al., 2013),  would still require a strong composition  
gradient between the H-He-rich outer envelope and the ice-rich interior in Neptune, and therefore the same argument 
of a TBL, if caused by the composition gradient, should also apply to Neptune. 
Its deeper possible location at $60\%R_{\rm Nep}$ is insufficient to explain Neptune's strong heat flow and the heat flow difference to Uranus. Under the same assumption of a monotonously growing TBL in Neptune, its cooling time would fall short as shown by the solution labeled [N/S16] in Figure \ref{figNN:taus}.

\begin{figure}[h]
\centering
\includegraphics[width=7.5cm]{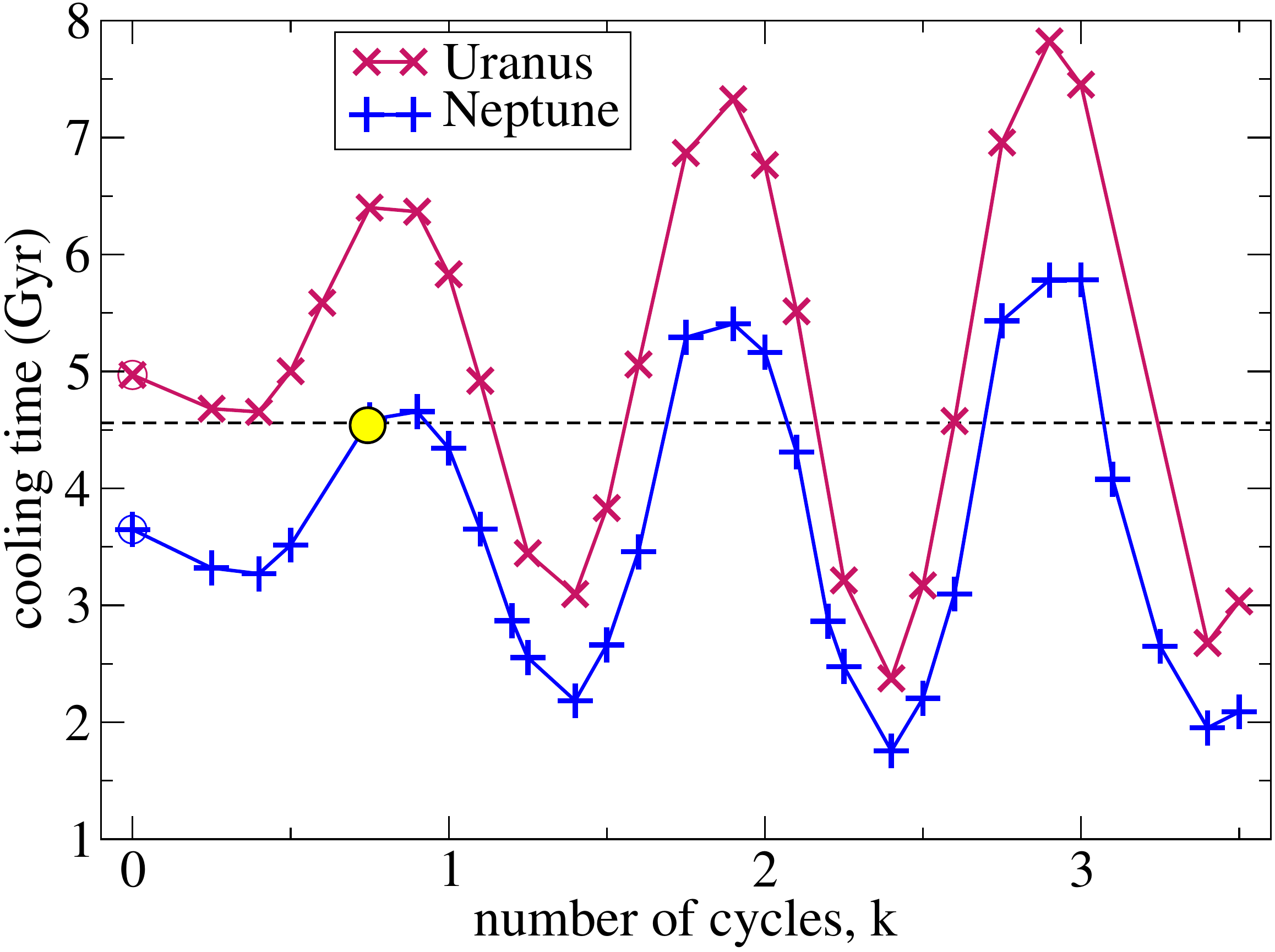}
\caption{\label{figNN:cycles}Cooling times of Uranus (red) and Neptune (blue) under the assumption of a slowly oscillating thermal boundary layer that builds up to maximum temperature difference of $\Delta T=1000\:$K 
(cycle number $k=n+1/2$ and completely decays again (cycle number $k=n$, $n$ natural number). The suggested 
current state of Saturn (Leconte \& Chabrier, 2013) is marked by a yellow circle. Thus after $n$ cycles, the thermal boundary layer (TBL) has been build up and decayed $n$ times.}
\end{figure}

If Uranus and Neptune have similar internal structures, how can the different heat flows be explained?  
If Neptune is not excessively rock-rich and its atmospheric  Albedo is not much lower than 
that of Uranus, recent evolution models find the planet to be excessively bright (Scheibe et al., 2019). Another
excessively bright planet is Saturn. One possibility to explain the brightness of Saturn is a thick thermal 
boundary layer that has retarded the loss of the intrinsic heat from the time of formation but allows it to 
slowly escape at present along a superadiabatic gradient (Leconte \& Chabrier, 2013, Vazan et al., 2016). Transferring this idea to 
Neptune implies that the temperature gradient across the TBL, if there is any, has already surpassed its 
maximum and is now decaying, releasing the primordial heat from the deep interior. In Figure \ref{figNN:cycles} 
we show the cooling times of Uranus and Neptune with a slowly oscillating temperature gradient $\Delta T/dr$ 
across the TBL of width $dr\ll R_p$. The TBL is assumed to adopt a maximum of $\Delta T=1000$ K before decaying. 
This process of a growing and decaying
TBL may occur repeatedly, leading to cycles in the temperature jump at the TBL. As a result, the cooling time of Uranus and Neptune
may shorten or prolong, depending on the state of the cycle. A significant shortening of the cooling time is found when the temperature jump
reaches its maximum, while faster  cooling times than the adiabatic case (cycle number $k=0$) are found when the TBL has decayed again.  
The current state of Saturn is shown by a yellow circle though in that case the TBL is proposed to undergo only 
one cycle, which extends to infinity (Leconte \& Chabrier, 2013).
While this is clearly a toy model, it illustrates the power of stably stratified layers and the importance of 
understanding  their heat transport efficiency and temporal evolution for understanding the internal structure 
and evolution of Uranus and Neptune. This toy model can explain the heat flow of both 
Uranus and Neptune.

\subsection{Non-Adiabatic Evolution Models}
As discussed above, both Uranus and Neptune are likely to have non-adiabatic deep interiors. 
It is therefore required to model the evolution (and internal structure) of the planets in a more realistic way in which the heat transport is calculated by the local conditions as time progresses.
The reason for a non-adiabatic planetary structure can be primordial composition gradients.  Such gradients can surpass convection and slow down the planetary cooling. 
In this case the internal structure can change in time by mixing of composition in convective regions. The change in the interior structure affects the planetary thermal evolution, and therefore should be considered self-consistently with the thermal evolution.
\par 

Recently, Vazan \& Helled (2019) calculated Uranus' evolution for various initial composition distribution profiles.
It was found that there are several types of composition gradients that fit Uranus low luminosity, as presented in Fig.~11. The deep interior of these models can be very hot, in spite of the planet's low luminosity. 
The existence of a stable composition gradient in Uranus also indicates that Uranus' current-state internal structure is not very different from its primordial one. Such a gradual structure also constrains the initial energy budget of the planet, and suggests that the initial energy content cannot be greater than 20\% of Uranus formation (accretion) energy, in order to fit the measurements.

\begin{figure}[h]
\centering
\includegraphics[width=12.5cm]{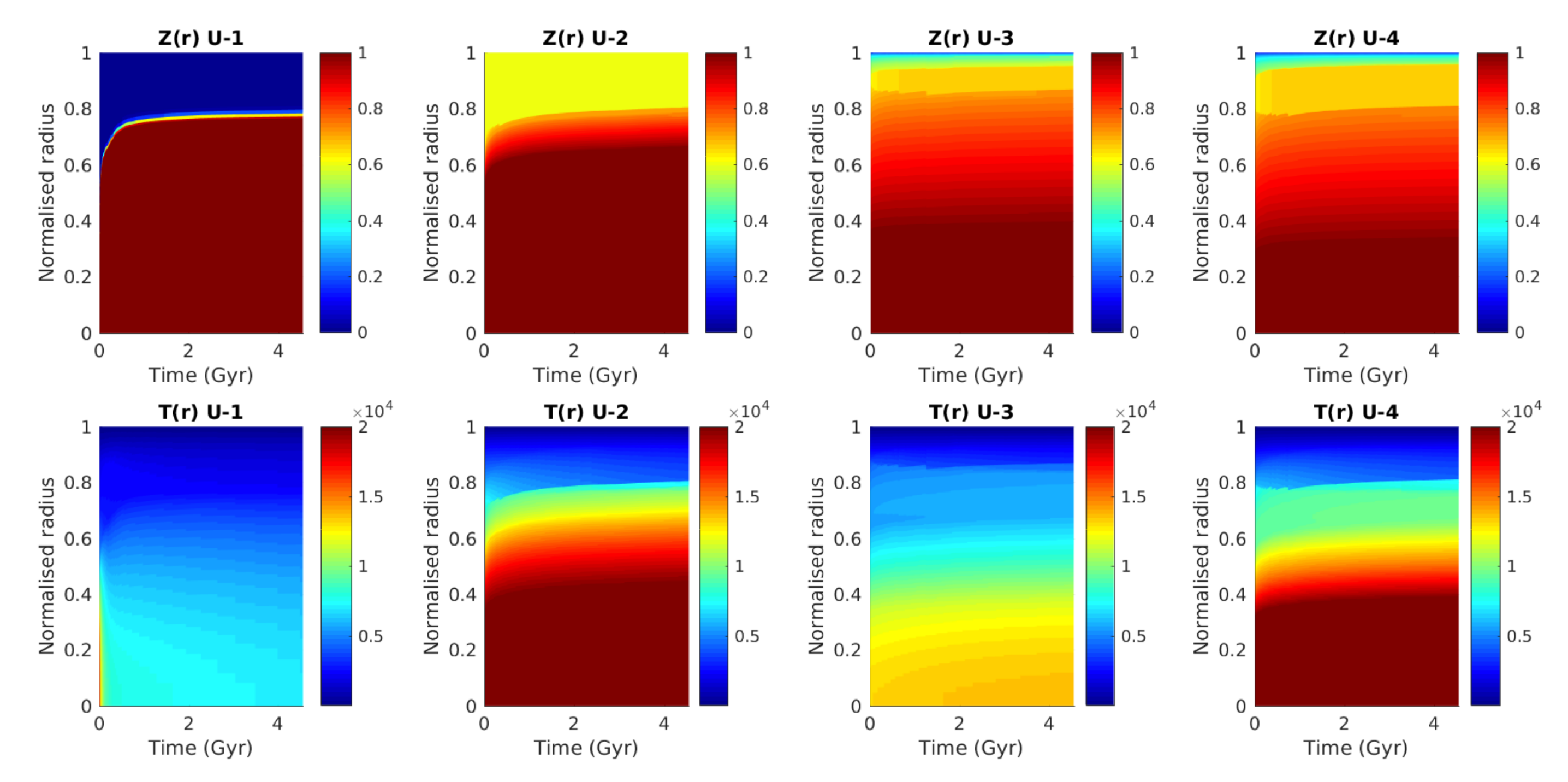}
\caption{Representative non-adiabatic models of Uranus evolution vs.~radius (y-axis) and time (x-axis). 
{\bf Top}: the heavy-element mass fraction $Z(r)$. {\bf Bottom}: the temperature profile $T(r)$. 
All the models are consistent with Uranus' observed parameters despite their different internal structures: 2-3 layer model (U-1), steep gradient model (U-2), a shallow composition gradient model (U-3), and a rock-rich composition gradient (U-4). Figure from Vazan \&Helled, 2019.}
\end{figure}

It was concluded that a composition gradient in Uranus' interior naturally explains its low luminosity, without the need of artificial thermal boundaries. Different types of composition gradients are stable during the evolution and are sufficient to slow down the cooling and fit the observed radius, moment of inertia, and luminosity.  
Interestingly, the total heavy-element mass fraction in Uranus is affected by the non-adiabatic evolution, and the hot gradual models result high metallicity for the planet (up to 95\%).
Such an evolution-interior path could also be relevant for Neptune. The fact that Neptune's luminosity seems to be consistent with adiabatic cooling does not necessarily mean that it is indeed adiabatic. This topic, which also reflects on our understanding of intermediate-size exoplanets, should be further investigated in future studies.  

\newpage
\section{Summary and future}
\label{sec:5}
Uranus and Neptune represent a unique class of planets in the Solar System, and yet, we know very little about them. 
As discussed in this review, Uranus and Neptune are mysterious planets in terms of their formation and evolution paths, and internal structures. 
While Uranus and Neptune clearly represent a distinct population of planets as they differ from heavy-element dominated terrestrial planets and the H-He dominated gas giants, it is still unclear how different the two planets are from each other. \\
Some key open questions are summarized below:
\begin{enumerate} 
{\it 
\item[-] How and where did Uranus and Neptune form?
\item[-] What is the bulk composition of the planets? Are the ice giants really water-rich?
\item[-] Are the planets mostly convective? Do they consist of boundary-layers/composition gradients?
\item[-] Where and how are the magnetic field generated? 
\item[-] Do the planets have water oceans? 
\item[-] What are the atmospheric compositions of Uranus and Neptune and how are they linked to the deep interior? 
\item[-] What are the causes for the observed differences between the two planets? Did the planets suffer from giant impacts?
}
\end{enumerate} 

Since the flybys of Voyager 2 near these planets, only more questions have been raised, putting Uranus and Neptune in the focus of planetary science studies. The goal to understand these planet  became even more profound with current statistics of exoplanets suggesting that planets in Uranus/Neptune masses and sizes are very common in our galaxy. 
It is therefore clear that a dedicated mission to these planets is highly desirable.  
Both NASA and ESA recognized the importance of Uranus and Neptune, and there are currently several mission proposals dedicated to the ice giants exploration. \\

We hope that these efforts will be successful and that a mission(s) to Uranus and Neptune will become reality. In particular, we suggest to measure their gravitational and magnetic fields, shapes, as well as their atmospheric properties, and the abundance of key elements and molecules. 
A better understanding of Uranus and Neptune will not only advance the fields of planetary science and astrophysics, but will also impact other fields such as space science, high-pressure physics, and geoscience. 
\par

\begin{acknowledgements}
We thank W.~B.~Hubbard and an anonymous referee for their valuable comments and careful reading of the manuscript.  
We also acknowledge some inspiring discussions within the ISSI "Formation of the Ice Giants" team meeting (Bern, 2019). 
\end{acknowledgements}

\clearpage


\end{document}